\documentclass[namedreferences,hyperref,optionalrh,solaromanenum]{spr-sola}
\usepackage{graphicx}
\usepackage{amssymb}  
\usepackage{amsmath}      
\usepackage{color}    
\usepackage{fixltx2e}       

\usepackage{textcomp}     
\usepackage{gensymb}
\usepackage{authblk}  
\usepackage{subfig}
\begin{document}
\begin{frontmatter}
\title{Visible Emission Line Coronagraph (VELC)  on board Aditya-L1}
\author[addressref=aff1]{\fnm{{Jagdev}~\snm{Singh}}}
\author[addressref=aff1,corref,email={ramesh@iiap.res.in}]{\inits{R.Ramesh}\fnm{R. Ramesh}}
\author[addressref=aff1]{\fnm{B. Raghavendra}~\snm{Prasad}}
\author[addressref=aff1]{\fnm{V. Muthu}~\snm{Priyal}}
\author[addressref=aff1]{\fnm{K. Sasikumar}~\snm{Raja}}
\author[addressref=aff1]{\fnm{S.N.}~\snm{Venkata}}
\author[addressref=aff1]{\fnm{P.U. Kamath}}
\author[addressref=aff1]{\fnm{V. Natarajan}}
\author[addressref=aff1]{\fnm{S. Pawankumar}}
\author[addressref=aff1]{\fnm{V.U. Sanal}~\snm{Krishnan}}
\author[addressref=aff1]{\fnm{P. Savarimuthu}}
\author[addressref=aff1]{\fnm{Shalabh}~\snm{Mishra}}
\author[addressref=aff1]{\fnm{Varun}~\snm{Kumar}}
\author[addressref=aff1]{\fnm{Chavali}~\snm{Sumana}}
\author[addressref=aff1]{\fnm{S. Bhavana}~\snm{Hegde}}
\author[addressref=aff1]{\fnm{D. Utkarsha}}
\author[addressref=aff1]{\fnm{Amit}~\snm{Kumar}}
\author[addressref=aff1]{\fnm{S. Nagabhushana}}
\author[addressref=aff1]{\fnm{S. Kathiravan}}
\author[addressref=aff1]{\fnm{P. Vemareddy}}
\author[addressref=aff1]{\fnm{C. Kathiravan}}
\author[addressref=aff1]{\fnm{K. Nagaraju}}
\author[addressref=aff1]{\fnm{Belur Ravindra}}
\author[addressref=aff1]{\fnm{Wageesh Mishra}}

\address[id=aff1]{Indian Institute of Astrophysics, Koramangala, Bengaluru - 560034}
\runningauthor{Singh et al.}
\runningtitle{\textit{ VELC on Board Aditya-L1}}



\begin{abstract}
Aditya-L1,  India's first dedicated mission to study the Sun and its atmosphere from the Sun-Earth Lagrangian L1 location was successfully launched on 2023 September 2. It carries seven payloads. The Visible Emission Line Coronagraph (VELC) is a major payload on  Aditya-L1. VELC  is designed to carry out imaging and spectroscopic observations (the latter in three emission lines of the corona), simultaneously. Images of the solar corona in the continuum at 5000\,{\AA}, with  a field of view (FoV) from 1.05\,$\mathrm{R}_\odot$ to 3\,$\mathrm{R}_\odot$ can be obtained at variable intervals depending on the data volume that can be downloaded. Spectroscopic observations of the solar corona in three emission lines, namely 5303\,{\AA}  Fe{\sc xiv}, 7892\,{\AA} Fe{\sc xi}, and 10747\,{\AA} Fe{\sc xiii} are possible simultaneously, with different exposure times and cadence. Four slits, each of width  50\,${\mu}$m, separated by 3.75\,mm help to simultaneously obtain spectra at four positions in the solar corona at all the aforementioned lines.  A Linear Scan Mechanism (LSM) makes it possible to scan the solar corona up to ${\pm}$1.5\,$\mathrm{R}_\odot$. The instrument has the facility to carry out spectropolarimetric observations at 10747\,{\AA} also in the FoV range 1.05\,-\,1.5\,$\mathrm{R}_\odot$. Various components of the instrument were tested interferometrically on the optical bench before installation. The individual components were aligned and performance of the payload was checked in the laboratory using a laser source and tungsten lamp. Wavelength calibration of the instrument was verified using  Sun as a light source. All the detectors were calibrated for different parameters such as dark current and its variation with exposure time. 
Here, we discuss the various features of the VELC, alignment, calibration, performance, possible observations, initial data analysis and results of initial tests conducted in-orbit.
\end{abstract}

\end{frontmatter}
\section{Introduction}
The atmosphere of the Sun is highly dynamic, where the variations occur with periods ranging from seconds to years. Further,  the physical nature of solar coronal structures are very complex. The energetic events occurring at the Sun  lead to solar flares and coronal mass ejections  that affect space weather and cause disturbances on the Earth. Sometimes during very energetic events and coronal mass ejections (CMEs), electrical grids and communication systems get damaged. To minimize the effect of such  events, the Sun and its atmosphere need to be monitored all the time, 24 hours a day throughout the year. Apart from this, the plasma in the solar corona gets heated up to millions of degrees as compared  to the Sun's surface temperature at 5700$^\circ$ K. There are various models to explain the heating of  the solar corona such as, wave heating or impulsive heating. The role of these processes need to be  understood. There are also contradictions in explaining the flows and/or waves in the solar corona. If there are waves and if these waves get damped, then, at what height  are they damped?. 

The invention of  the coronagraph by Lyot in 1930 enabled observations of the extended coronal atmosphere of the Sun by blocking the solar disk. This invention opened the doors to study the solar corona without the occurrence  of a total solar eclipse, with ground based coronagraphs as well as space based instruments. Various space missions such as  the SOlar and Heliospheric Observatory (SOHO),  the Solar Dynamics Observatory (SDO),  the Solar Terrestrial Relations Observatory (STEREO),  the Transition Region  and Coronal Explorer (TRACE), Ulysses, Hinode, Solar Orbiter (SO), the Parker Solar Probe (PSP), and many others have observed the Sun and its atmosphere in different wavelength domains such as X-rays, ultraviolet (UV) and extreme-ultraviolet (EUV). We had planned the VELC for continuous monitoring of the Sun everyday for 24 hours considering the following results reported from observations during the total solar eclipses, and with the ground-based coronagraphs.  It is well known that the duration of observations in both the cases are limited due to obvious reasons.

Most of the spectroscopic or imaging data in the visible wavelength range have been recorded during total solar eclipses or using coronagraphs during excellent clear sky conditions
(\citealp{pasachoff2002,raju2011,voulgaris2012,koutchmy1983,ichimoto1995,singh1982, singh2003a,singh2003b,singh2004a,singh2004b,singh2006,singh2011a,samanta2016,rusin1994, minarovjech2003}). The observations during total solar eclipses provide data with minimum scattering by the Earth’s atmosphere. The advantage of coronagraphs in the visible wavelength is that information with height from the solar limb becomes available for relatively longer  durations (\citealp{singh2004b}).
Spectroscopic observations in the transition region spectral lines obtained in the UV and EUV wavelength bands provide information up to about 
1.2\,$\mathrm{R}_\odot$ (\citealp{hassler1990, doyle1998}). These correspond to plasma at relatively low temperatures compared to that of  the coronal plasma. In the X-ray wavelength domain, most of the observations have been made either by imaging the corona to study  the dynamics of coronal structures as function of height (\citealp{krishna2013}) or by performing spectroscopy of coronal loops on the solar disk to investigate their physical and dynamical properties as a function of distance from the base to the  top of the loop (\citealp{krishna2017}). This way the true height from the solar limb is not known.  Furthermore, the signal gets integrated with the background solar disk and height along the loop. Spectroscopic observations of energetic events have been made in the emission lines in the X-ray wavelength range during the occurrence of jets which propagate outwards from the solar limb (\citealp{jelinek2015}). The oscillations in the Doppler velocity with a period of ${\approx}$300\,sec were detected by \cite{tsubaki1977} at some locations along the slit while observing in 5303\,{\AA} emission line. \cite{koutchmy1983} found velocity oscillations with periods near 300\,sec, 
80\,sec and 43\,sec from the time sequence spectra in the 5303\,{\AA} line at 1.04\,$\mathrm{R}_\odot$. But they did not find any prominent intensity fluctuations during the period of observations. \cite{singh1997} found intensity oscillations in the solar corona having periods of 56.5\,sec, 
19.5\,sec, 13.5\,sec, 8.0\,sec, 6.1\,sec, and 5.3\,sec with amplitudes in the range of 0.2\,-\,1.3\% of the coronal brightness.

\cite{singh1999,singh2003a,singh2003b,singh2004b,singh2006} made systematic observations of the solar corona in four emission lines simultaneously during the period 1997\,-\,2007 with the 25\,cm coronagraph at Norikura observatory, Japan. Most of the time they could observe the solar corona between 
1.01\,$\mathrm{R}_\odot$ and 1.20\,$\mathrm{R}_\odot$ in the 5303\,{\AA}, 
6374\,{\AA}, 7892\,{\AA}, 10747\,{\AA} and 10798\,{\AA} coronal emission lines.
They could carry out spectroscopic observations up to 1.5\,$\mathrm{R}_\odot$  during a couple of days only because of limited number of hours of clear coronagraphic skies. \cite{raju2011} found that the width of the 5303\,{\AA} emission line does not vary with height above the solar limb using data obtained with the Fabry-Perot technique  
during the total solar eclipse of  2001 June 21, from Lusaka. But \cite{singh1999, singh2003a, singh2003b, singh2004b, singh2006}, and \cite{krishna2013} found variations in all the emission lines with height above the limb in  a large number of observations made during the period  1997\,-\,2007 with the 25\,cm coronagraph mentioned above. The observations in different emission lines obtained on different days and of different coronal regions indicate  a similar behavior irrespective of shape, size and direction in the sky plane. They found  a decrease in the line width with height in the Fe{\sc xiv} line and an increase in the line width with height in 
Fe{\sc x} (6374\,{\AA}) and Fe{\sc xi} emission lines. The Fe{\sc xiii} emission line width shows small increase or decrease with height. The magnitude  of the increase/decrease with height was different in different coronal structures, up to about 1.20\,$\mathrm{R}_\odot$ in all the observations. 
 The reason for not detecting the decrease in line-width of the green line with height by \cite{raju2011}) may be because of the lower spectral resolution of 26,000 in their Fabry-Perot interferometer observations as compared to the higher spectral resolution of 4,80,000 in the spectroscopic observations with the 25\,cm Norikura coronagraph mentioned above. \cite{singh2006} could obtain a raster scan of spectra in the Fe{\sc x} and Fe{\sc xiv} emission lines with long exposure time for each spectra from 1.01 to 1.5\,$\mathrm{R}_\odot$ on 2003 October 26 due to availability of long hours of excellent clear skies. They found that line widths of Fe{\sc x} increase whereas  those of Fe{\sc xiv} decrease with height above the limb. The increase and decrease of line width happens up to about 200$^{\prime\prime}$ above the limb, but remains the same beyond this height. 

Further, these observations indicate that  the intensity ratio of Fe{\sc xi}$/$Fe{\sc x} increases with height above the limb whereas that of Fe{\sc xiv}$/$Fe{\sc x} and Fe{\sc xiv}$/$Fe{\sc xiii} decreases with height above the limb (\citealp{singh2004b}; \citealp{krishna2013}). The systematic observations show that variations in the intensity ratios and line widths for different lines are complex. Similarly, the line width ratio Fe{\sc x}$/$Fe{\sc xi} increases whereas that of Fe{\sc xiv}$/$Fe{\sc x} decreases with height above the limb. These results imply that loop tops are hotter as compared to the  footpoints if observed in Fe{\sc x} and Fe{\sc xi} emission lines, and loop tops are cooler if observed in Fe{\sc xiv} and Fe{\sc xiii} emission lines. This type of phenomena is difficult to visualize and explain. Such type of variation in the parameters of emission lines is contradictory, considering the abundances of these ions as function of temperature. It may be noted that these results are based on the observations up to  
${\approx}$1.2\,$\mathrm{R}_\odot$ only. It is still not known how the intensity ratios vary beyond 1.2\,$\mathrm{R}_\odot$. Such type of behavior of the coronal emission lines needs to be confirmed and investigated further by carrying out systematic observations. It may also be noted that  the spectral resolution of observations made in the visible part of the spectrum is much larger than that in UV, EUV and X-rays.

In view of the projected science objectives to study the physical and dynamical  structure of  the solar corona, space weather by monitoring the occurrence and dynamics of CMEs, magnetic topology  of  the solar corona, and address the heating mechanism  of the coronal plasma, VELC was developed (\citealp{singh2011b}). We have planned  the imaging of the solar corona at 5000\,{\AA} over the  FoV 1.05\,-\,3\,$\mathrm{R}_\odot$, and spectroscopy around three emission lines (5303\,{\AA}, 7892\,{\AA}, 
10747\,{\AA}) over the FoV 1.05\,-\,1.5\,$\mathrm{R}_\odot$, simultaneously. We have listed the wavelength values of emission lines based on the ground measurements, as most of the readers are familiar with the wavelengths of these emission lines. The filters have been designed taking into account their use in space (vacuum conditions) and the convergence of beam passing through the filter. Here, we discuss the requirements of the instrument, design, calibration, and present status of the payload. 

\section{Requirements and Purpose of VELC}

Observations of the solar corona during total solar eclipses can be carried out for few minutes only from any location along the totality path of the eclipse.
Ground based observations using a coronagraph require excellent sky conditions which generally are limited to a couple of hours and that too in select locations only for minimal number of days in a year. Coronal observations in the visible emission lines provide better spectral resolution as compared to those in EUV and X-ray wavelengths. The Large Angle and Spectrometric Coronagraph (LASCO) C1 on board SOHO had the facility to observe the corona in visible emission line using Fabry-Perot etalon, but with comparatively 
 low spectral resolution. To monitor the solar coronal energetic events 
24 hours a day throughout the year, the requirement of a space coronagraph operating at visible wavelengths with better spectral resolution was realized. The Indian scientific community after a large number of meetings decided to have a space coronagraph with the following science objectives:
\begin{itemize}
    \item Diagnostics of the coronal parameters (temperature, velocity, and density).
    \item To understand the processes involved in heating of the corona and solar wind acceleration.
    \item To study the origin, development, and dynamics of CMEs.
    \item To study the dynamics of the large scale coronal transients.
    \item To study the drivers of space weather.
    \item To measure the coronal magnetic fields.
\end{itemize}

An internally occulted coronagraph was designed to carry out the above mentioned scientific objectives. To study the physical and dynamical characteristics of the coronal plasma and features such as coronal loops, spectroscopy in two or more emission lines is needed. Three emission lines were chosen to cover a temperature range of 1\,-\,2 ${\times}10^{6}$  K. It is desirable to carry out observations as close to the solar limb as possible. Considering the pointing limitations of the satellite and scattering due to  the primary mirror, it was decided to have observations in the heliocentric distance ($r$) range 1.05\,-\,3.0\,$\mathrm{R}_\odot$ for imaging, and 1.05\,-\,1.5\,$\mathrm{R}_\odot$ for spectroscopy. Typically, observations with spatial resolution 
${\approx}$ 1\,-\,2\,$^{\prime\prime}$, spectral resolution 
${\sim}$ 10\,-\,100\,m{\AA}, and temporal resolution 
${\sim}$ 1\,-\,100\,sec  are preferable to study the `quiet' corona and energetic events. A two-beam arrangement was planned to perform spectro-polarimetric observations also. After preliminary optical design, feasibility studies were carried out with the vendors and various technical divisions of Indian Space Research Organisation (ISRO). The discussions indicated that the primary mirror of the coronagraph needed to be 200\,mm instead of 300\,mm as proposed initially. The size and weight of the payload imposed limitations. Three emission lines were chosen instead of the four emission lines proposed. The focal length of the primary mirror and spectrograph lens were also reduced.  The download limit of 100 Gigabit per day data restricted the provision to carry out high frequency (1\,Hz) observations. Various parameters of the final design are listed in Table \ref{tab:1}. Photon computations and other details are described in \cite{kumar2018} and \cite{singh2019}.

\begin{table}[h!]
  \caption{Parameters of the internally occulted mirror coronagraph}
    \label{tab:1}
    \begin{tabular}{ll}
        \hline
        Parameter &  Specification \\ \hline
        Imaging of solar corona   & At 5000\,{\AA} with FoV up to 
        3\,$\rm R_{\odot}$     \\ 
        Spectroscopy of solar corona  & FoV up to 
        1.5\,$\mathrm{R}_\odot$     \\ 
        Emission lines for spectroscopy & 5303\,{\AA} Fe{\sc xiv}, 
        7892\,{\AA} Fe{\sc xi}, \\
         & 10747\,{\AA} Fe{\sc xiii}    \\ 
        Primary mirror  & Parabolic 195\,mm diameter,  \\
       & focal length = 1300\,mm, f-ratio\,=\,6.67    \\ 
        Entrance Aperture  & 147\,mm    \\ 
        Littrow lens focal length  & 780\,mm     \\ 
        Multi-slit & 4 slits, each of width 50\,${\mu}$m 
        (9.6$^{\prime\prime}$), \\ 
        & spacing between the slits\,=\,3.75\,mm \\ 
        Grating & 600 lines/mm blazed at 42$^{\circ}$ \\ 
        Pixel size of visible detector (sCMOS) & 6.5\,${\mu}$m$^{2}$ \\ 
        Pixel size of IR detector (InGaAs) & 25\,${\mu}$m$^{2}$ \\ 
        Plate scale & 
        2.25$^{\prime\prime}$/pixel (Continuum) \\ 
        & 1.25$^{\prime\prime}$/pixel (5303\,{\AA} \& 7892\,{\AA})  \\ 
        & 4.33$^{\prime\prime}$/pixel (10747\,{\AA})  \\ 
        Spatial resolution & 4.5$^{\prime\prime}$ (continuum) \\
        & 2.5$^{\prime\prime}$ (5303\,{\AA} \& 7892\,{\AA})  \\   
        & 8.6$^{\prime\prime}$ (10747\,{\AA})  \\ 
        Spectral Dispersion & 28.4\,m{\AA}/pixel (Fe{\sc xiv}, 4th order) \\
        & 31.3\,m{\AA}/pixel (Fe{\sc xi}, 3rd order) \\
        & 223.0\,m{\AA}/pixel (Fe{\sc xiii}, 2nd order) \\ 
        Efficiency & ${\approx}$28\% (continuum channel) \\
        & ${\approx}$5\% (spectral channels) \\ 
        \hline
    \end{tabular}
      \end{table}

\section{Optical Design of VELC}
Considering the science objectives mentioned in Section 2, the instrument should have facility to image the solar corona in continuum, perform spectroscopy in coronal emission lines, and spectropolarimetry to measure the polarization of solar corona. We planned to have these in a single payload by imaging the solar corona at 500\,nm, and carry out spectroscopy in emission lines at wavelength $>$500\,nm by splitting the coronal beam using a dichroic beam splitter. Based on the experience of observations with the 25\,cm coronagraph at Norikura observatory in Japan, we chose the emission lines for VELC. It has the facility to observe in three emission lines 5303\,{\AA}, 7892\,{\AA}, and 10747\,{\AA}, simultaneously. 
The facility to observe in the 6374\,{\AA} emission line could not be included because of increase in width and weight. Simultaneous measurements of the line profiles in different emission lines are needed to derive the thermal and dynamic characteristics of coronal loops and structures. So we have used four detectors, one for the imaging in continuum and three for the above mentioned emission lines.

Data from observations close to the solar limb are very important to study the dynamics of CMEs and other coronal transients (see, e.g. \citealp{ramesh1999,kathiravan2002,ramesh2003,kishore2015,kumari2019}). Most of the coronal structures in emission do not extend to large distances in the solar corona and therefore require observations close to the solar limb to investigate their physical and dynamical characteristics \citep{ramesh2006}. In view of the pointing accuracy and drift of the spacecraft in space, it was decided to have a FoV from 1.05\,$\mathrm{R}_\odot$. The available detector size, and the need for high resolution in the spectral observations restricted the outer FoV to 1.5\,$\mathrm{R}_\odot$ and 3.0\,$\mathrm{R}_\odot$ for the spectroscopic and continuum observations, respectively. Due to the above mentioned  inner limit of the FoV at 1.05\,$\mathrm{R}_\odot$, it was decided to develop an internally occulted coronagraph. In view of the spacecraft platform size, limit on the payload weight, the VELC instrument was designed with the details mentioned in the following paragraphs.
 
\begin{figure}[ht]
  \centering
  \includegraphics[width=0.99\textwidth,clip=]{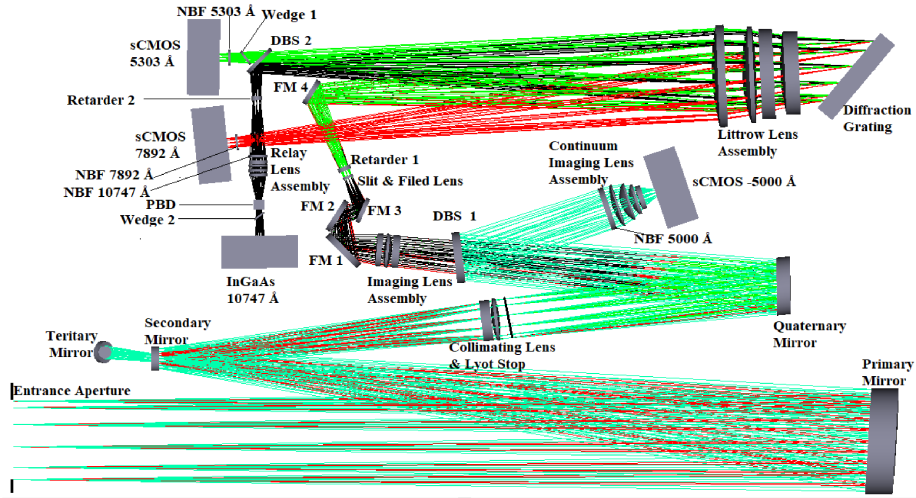} 
  \caption{Optical Layout of the Visible Emission Line Coronagraph (VELC) 
  on board Aditya-L1.}
  \label{fig:1}
\end{figure}

Figure~\ref{fig:1} shows the optical layout of VELC on board Aditya-L1 mission. The details of the optical design and its performance are given in \cite{kumar2018}, \cite{prasad2017}, and \cite{singh2019}. The off-axis concave parabolic mirror (M1) of VELC with a focal length 1300\,mm has a clear aperture of 192\,mm. The mirror was cut from the parent parabola at an off-axis  distance of 152\,mm. M1 is mounted on the optical bench at a distance of 1570\,mm from entrance aperture (EA) of size 147\,mm to cover the desired FoV of 3\,$\mathrm{R}_\odot$. The diameter of the solar disk beam is 162\,mm on the primary mirror. For the coronal beam corresponding to the 3\,$\mathrm{R}_\odot$ FoV, it will be 192\,mm. The Secondary Mirror (M2), concave and spherical with central elliptical hole is positioned at the focal plane of M1. A Tertiary mirror (M3) installed behind M2 at an angle of 45$^{\circ}$ reflects the light up to 0.28$^{\circ}$ 
(1.05\,$\mathrm{R}_\odot$) into deep space through a small opening on the top cover of the payload. The small opening is made such a way that outside light does not enter into the instrument. The optical surface of M2 reflects the solar coronal light over the FoV 
1.05\,-\,3\,$\mathrm{R}_\odot$ towards the Collimator Lens Assembly (CLA), a two-lens system designed to re-image the EA in combination with M1 and M2, on the Lyot-stop. It is kept at the pupil plane  of the CLA and stops the diffracted light from the EA proceeding to the quaternary mirror (M4) mirror. The CLA collimates the coronal light over the FoV 
(1.05\,-\,3\,$\mathrm{R}_\odot$) and  directs it towards M4. The latter is an off-axis concave parabolic mirror with  focal length of 1180\,mm, 
 off-axis distance of 250\,mm and clear aperture 106\,mm. It directs the coronal light towards the Dichroic Beam Splitter-1 (DBS-1) which reflects the coronal light $<$5150\,{\AA} towards  the continuum channel for imaging the corona and transmits the coronal light $>$5150\,{\AA} for spectroscopy observations. A four-element Imaging Lens Assembly-1 (ILA-1) forms the image of the solar corona with the light reflected from DBS-1. A narrow band filter with bandwidth of 10\,{\AA} centred at 5000\,{\AA} placed 
in  front of the lens assembly restricts the bandwidth of the continuum image  of the solar corona. Continuum image data are recorded by using a sCMOS camera of 2592\,${\times}$\,2192 format with pixel size 6.5\,${\mu}$m$^{2}$.  The plate scale of the continuum channel is 2.25$^{\prime\prime}$ per pixel. The transmitted coronal light with wavelength $>$5150\,{\AA} is focused by the Imaging Lens Assembly-2 (ILA-2). The latter is a  
three-element lens and it forms a color corrected image of the solar corona over a spectral band of 5150\,-\,11000\,{\AA} on the multi-slit assembly (MSA) of the spectrograph. A linear scan mechanism, consisting of two fold mirrors (FM-1 and FM-2) located between ILA-2 and the slit plane, is used to scan the FoV across the slits. Another fold mirror FM3 directs the beam towards the MSA of the spectrograph. 

Fold Mirror-4 (FM-4) guides the beam towards a Littrow Lens Assembly (LLA) consisting of four elements which act as collimator and camera lens. A diffraction grating is used to disperse the incident beam. The size of the LLA has been designed such that the diffracted beam for all the three wavelengths (i.e. 5303\,{\AA}, 7892\,{\AA}, and 10747\,{\AA}) do not have any intensity vignetting. The Littrow lens directly focuses the spectra  around the 7892\,{\AA} (3rd order) emission line on the sCMOS detector.  A narrow band filter (NBF) centred at 7892\,{\AA} with 8\,{\AA} pass band kept in front of the detector avoids the overlap of spectra due to the use of multiple (four) slits. The second order line at 10747\,{\AA} and the fourth order line at 5303\,{\AA} lie almost in the same direction. So, a Dichroic Beam Splitter-2 (DBS-2) with a small wedge to correct the vignetting for 5303\,{\AA} spectra, is used to separate the two dispersed beams spectrally and spatially. DBS-2 transmits the spectral band around 5303\,{\AA} and reflects the band around 10747 {\AA}. The transmitted spectra around 5303\,{\AA} gets focused through a 6\,{\AA} narrow band filter (NBF) which avoids the overlap of spectra due to the use of multiple slits. A rotating polarization modulator (quarter wave plate, retarder 2 in Figure \ref{fig:1}) is used to modulate the 10747\,{\AA} light beam in order to measure the Stokes parameters I,\,Q,\,U,\,\&\,V.  The details about the polarization measurements are  described in \cite{nagaraju2021,venkata2022,sasikumar2022,venkata2024}.

The reflected beam of 10747\,{\AA} comes to focus in between the focal planes of 5300\,{\AA} and 7892\,{\AA} (Figure \ref{fig:1}). Accommodating the detector
and the entire spectropolarimetry package in the space available between DBS-2 and 10747\,{\AA} focal plane is not possible. In order to overcome this issue, the image plane of the 10747\,{\AA} is relayed to a different location by using a relay lens assembly with unit magnification (\citealp{venkata2022}). A Polarizing Beam Displacer (PBD) assembly along with a wedge is mounted after the RLA to split the E- and O-rays to perform the polarization measurements. This arrangement permits the spectroscopic as well as polarization observations in the 10747\,{\AA} emission line. The retarder remains stationary during spectroscopy and rotates during polarization observation. The retarder completes one rotation in 10.08 seconds. More details about the PBD, RLA, and retarder are described in \cite{sasikumar2022}.

The payload was assembled, aligned, and tested for its performance in class 10 area of the laboratory to minimize molecular and particle deposition on the optical components of the instrument. Only two persons worked inside the laboratory at any given time to maintain its class 10 quality. In the following sections, we give details of some components of the payload. 

\subsection{Shutter with the Neutral Density (ND) Filter}
A multi-time opening shutter has been mounted in front of the EA of the coronagraph. This shutter will be closed during the operation of spacecraft thrusters to avoid the contaminants from entering the instrument. The shutter has a neutral density (ND) filter of 50\,mm diameter and density 4, calibrated at the required wavelengths. The ND filter is of reflector type to minimize its heating due to sunlight. The transmission of the filter is 
$0.56{\times}10^{-4}$, $0.64{\times}10^{-4}$, $1.51{\times}10^{-4}$, and $2.95{\times}10^{-4}$ at 5000\,{\AA}, 5303\,{\AA}, 7892\.{\AA}, and 
10747\,{\AA}, respectively. The clear aperture of the ND filter is 45\,mm. The density of the ND filter, and ratio of the sizes of the EA and ND filter limits the intensity of the solar disk light in the shutter closed position to nearly the same as that of the coronal light in shutter open condition. The purpose of the ND filter is to obtain images and spectra of solar disk by 
off-pointing the spacecraft in the shutter closed condition. This data will be used to convert the observed detector counts from the corona (during shutter open observations) in units of solar flux, i.e. 
ergs\,s$^{-1}$cm$^{-2}${\AA}$^{-1}$sr$^{-1}$.  

\subsection{Primary Mirror M1}
The M1 mirror parent parabola was fabricated and super polished at the Laboratory for Electro-Optics Systems (LEOS), an ISRO centre. Four mirrors were cut with an off-axis distance of 152\,mm. One of the off-axis mirrors with better quality was used in the flight model of the payload. 
To estimate the scatter from the mirror, measurements of micro-roughness were carried out in the laboratory using a profilometer for various plate scales in the range of 
20\,${\mu}$m to 2.15\,mm. The measured micro-roughness is 
${\approx}$5\,{\AA} for the plate scale of 2.15\,mm. The details about the micro-roughness and scatter measurements are given in \cite{venkata2017}. 

\subsection{M2 mirror as an Occultor and Reflector of Coronal Light}
Stainless steel was selected for the fabrication of the M2 mirror to obtain the sharp edge of the central hole, and coated with nickel using electrolysis process. The major and minor axes of the central hole of M2 are 13.05\,mm and 12.95\,mm, respectively. Clear aperture of the secondary mirror M2 is 40\,mm. The focal length is 700\,mm. The elliptical hole at the centre of M2 mirror acts as an internal occulter. Thus, the M2 mirror allows the solar disk light and coronal light up to 1.05\,$\mathrm{R}_\odot$ pass through it, and reflects the coronal light from 1.05 to 3.0\,$\mathrm{R}_\odot$ for imaging and spectroscopic observations.

\subsection{Multi-slit Assembly}
The multi-slit assembly (MSA) consisting of four slits was used to decrease the observing time by a factor of four for the spectroscopic observations over the required FoV. The MSA has four slits with separation of 3.75\,mm between each slit. Initially the width of the slits was proposed to be 20\,${\mu}$m to achieve a high spatial resolution 
of 3.8$^{\prime\prime}$. But the width of the slits was increased to 
50\,${\mu}$m considering the availability of photons in the emission lines in the outer corona around 1.3\,$\mathrm{R}_\odot$, the drift of the spacecraft 0.2$^{\prime\prime}$ per second, and the exposure time required to get sufficient signal-to-noise ratio (SNR) for emission line profiles \citep{singh2019}. The slit width has an accuracy of ${\pm}$1\,${\mu}$m. The separation and parallelism between the slits are accurate within 
${\pm}$5\,${\mu}$m.  The increase in slit width decreases the spatial resolution to 9.6$^{\prime\prime}$, but sufficient for the emission line corona. Moreover, keeping 20\,${\mu}$m slit width would not have helped in view of the drift rate, pointing accuracy, and the large exposure time required for the spectroscopic observations. The separation of 3.75\,mm between the slits has been optimized considering the dispersion of the spectra and the 
full width at half-maximum (FWHM) of the narrow band filters. 

\subsection{Linear Scan Mechanism (LSM)}

\begin{figure}[ht]
  \centering
  \includegraphics[width=0.99\textwidth,clip=]{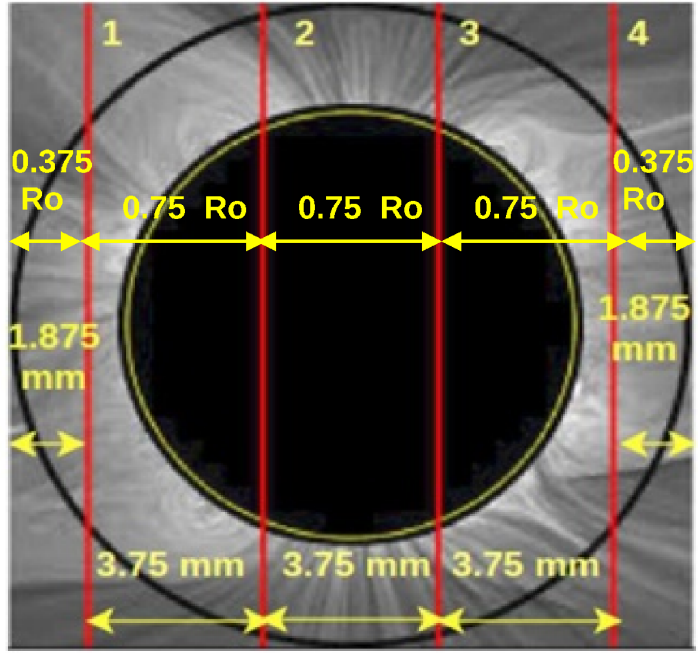} 
  \caption{Location of the four slits in the VELC spectroscopy channel. The vertical red lines represent the four slits and their locations on a coronal image (obtained during a total solar eclipse) when the LSM is in its `home' position. The separation between the slits is indicated by the horizontal yellow lines. The yellow circle at the center of the image represent the Sun disk size. The `filled' black circular area indicates the extent of the `hole' in the M2 mirror. The outer black circle indicates the FoV of VELC for spectrographic observations.}
  \label{fig:2}
\end{figure}

To obtain images of the solar corona in the three emission lines mentioned above, we need to scan the image across the four slits. The separation between the slits is 3.75\,mm. 
The image can be scanned across each slit by ${\pm}$1.875\,mm.
The image shifts by 20 ${\mu}$m at the slit plane when the LSM moves 
by 10\,${\mu}$m. Therefore, we need 
${\pm}$0.9375\,mm movement  of the LSM. For this purpose  the LSM was designed and developed at U R Rao Satellite Centre (URSC) of ISRO. The requirement was a motion  in steps of 10 ${\mu}$m step size and, traverse of 0.9375\,mm on either side of the middle position of the LSM termed as `home'. Movement of the LSM is read by an optical encoder. Two encoders are implemented (main and redundant). The LSM is designed for a travel range 
of ${\pm}$1.2\,mm (i.e. 
${\pm}$1200\,${\mu}$m). Considering possible inadvertent motion beyond 
${\pm}$0.9375\,mm, the LSM has a hardware stop switch at ${\pm}$1.2\,mm. Before this limit, there is a software stop at 
${\pm}$1\,mm. The location of the four slits on the coronal image at `home' position of the LSM is shown in Figure~\ref{fig:2}.   Two fold/flat mirrors, FM1 and FM2, mounted on a single platform are at 90$^{\circ}$ from each other, with FM1 perpendicular to the incoming beam. This platform is mounted on the LSM which can be moved in a direction joining the centre of these two mirrors. The movement of the LSM was tested in steps of 
10\,${\mu}$m and its multiples by measuring the movement of the image of a laser beam on a detector. The pixel size in the latter is 1.85\,${\mu}$m, for accurate measurements. The measurements show a deviation of ${\pm}$1\,${\mu}$m for a step size of 10\,${\mu}$m. 

\subsection{Dispersion Grating}

We planned to carry out spectroscopic observations simultaneously around  the 2nd order 10747\,{\AA}, 3rd order 7892\,{\AA} and 4th order 5303\,{\AA} emission lines, with maximum possible efficiency at all the wavelengths. A grating with 600 grooves/mm, blazed at 42$^{\circ}$, and of 
140\,mm${\times}$180\,mm size was required. We procured five replica gratings of the master grating fabricated with the specification mentioned above and used the first replica grating in the payload. With the optical set up of the instrument and this grating, the dispersion values are 28\,m{\AA}, 31\,m{\AA} and 223\,m{\AA} per pixel for 5303\,{\AA}, 7892\,{\AA} and 10747\,{\AA} emission lines, respectively.

\subsection{Filters}
As mentioned before, multi-slit assembly with four slits are used in VELC to decrease the observing time for generating images of the solar corona in the three emission lines. We had also planned to observe simultaneously in different wavelengths. Therefore we need to use narrowband filters to avoid overlap of spectra due to the different slits, and Dichroic beam splitters to separate the required wavelengths for each channel. We designed and then procured space qualified filters from Alluxa, Inc. The narrow band filters centred at 5304.3\,{\AA}, 7893.9\,{\AA} and 10749.3\,{\AA} (wavelengths in space) have bandwidths 
of 6\,{\AA}, 8\,{\AA}, and 11\,{\AA}, respectively. Their use in space and the shift in the passband of the filter in the converging beam were taken into consideration while specifying the central wavelength of the filters. 


\subsection{Detectors}

We have used four detectors, three sCMOS based detectors for the visible wavelength observations and one InGaAs detector (henceforth called IR detector) for the 10747\,{\AA} emission line. All the four detectors were designed, developed, and tested by Space Applications Center of ISRO. The performance of the chips was tested for the noise level at different operating temperatures. The noise decreases with operating temperature. 
In view of the noise and thermal design, it was decided to operate the sCMOS detectors at –5$^\circ$C and IR detector at –17$^\circ$C. The details of detector calibration and performance are given in \cite{singh2022},  \cite{priyal2023}. Some of the basic features of the detector are listed here: the sCMOS detectors have a 2592\,${\times}$\,2192 format with a pixel size of 
6.5\,${\mu}$m$^{2}$. It has the facility  to record the data in low and high gain, simultaneously. The low gains are 1X and 2X, high gains are 10X and 30X. Data can be obtained simultaneously in one low gain and one high gain. The full well capacity (FWC) is 30,000 electrons at 1X with 11-bit readout for both the gains. The IR camera has a 640\,${\times}$\,512 format with a pixel size of 
25\,${\mu}$m$^{2}$. It can be operated in two modes, one low gain and other high gain, but only one at a time. The readout is 12-bit. 

\section{Alignment and Integration of the Payload}

Based on the tolerance and performance analysis, fabrication specifications for each optical element and optical sub-assemblies were generated. Individual optical elements were tested and their performance was evaluated prior to integrating them as the sub-assemblies. Post realization of the sub-assemblies, they were subjected to the environmental tests and their performance was evaluated (\citealp{venkata2023a}; \citeyear{venkata2023b}). After completion of the testing and evaluation of the individual subsystems, they were aligned and integrated in the payload as per the 
pre-defined tolerances generated from the system level tolerance analysis. Alignment tolerances of optical sub-assemblies of VELC are listed in Table \ref{tab:2}. The entire process of alignment and integration of VELC is divided into seven major parts as mentioned below:
As mentioned in Table \ref{tab:2}, primary mirror (M1), collimating lens, and detectors are used as the compensators during the alignment process. The positions of these elements are adjusted in order to correct the aberrations/alignment errors of other optical elements, to meet performance requirements. 

\begin{table}[h!]
\caption{Alignment tolerances of optical sub-assemblies of VELC}
    \label{tab:2}
        \begin{tabular}{lccc}
        \hline
        
Optical & De-center & Tilt & Inter-separation \\
Sub-assembly & Tolerance  & Tolerance  &  Tolerance  \\ 
& (${\pm}{\mu}$m) & (${\pm}$arcsec) & (${\pm}{\mu}$m) \\
\hline
Primary mirror (M1) & Compensator & Compensator & Compensator \\
Secondary mirror (M2) & 50 & 30 & 50 \\ 
Tertiary mirror (M3) & 50 & 30 & 100 \\ 
Quaternary mirror (M4) & 50	& 30 & 100 \\ 
Collimating lens & Compensator & Compensator & Compensator \\ 
Continuum imaging lens & 50	& 30 & 100 \\ 
Spectrograph imaging lens & 50 & 30	& 100 \\ 
Littrow lens & 50 & 30 & 250 \\ 
Relay lens & 50	& 30 & 100 \\
Diffraction grating	& 100 & 30 & 100 \\ 
Dichroic beam splitters  & 100 & 30	& 100 \\ 
(DBS-1 and DBS-2) & & & \\
Narrow band filters & 100 & 30 & 100 \\ 
Fold mirrors & 100 & 30 & 100 \\ 
(FM1, FM2, FM3, and FM4) & & & \\
Detectors & Compensator & Compensator & Compensator \\ 
\hline
\end{tabular}
 \end{table}

\subsection{Optical Bench Alignment}

All the optical sub-assemblies of VELC are supported on a light weight optical bench made of titanium, mounted on the satellite top deck. The surface area on the top side of the optical bench for mounting the sub-assemblies has a surface planarity/irregularity $<{\pm}$20\,${\mu}$m. The optical bench was mounted on a transfer trolley which is fixed to a vibration isolation table in ISO-4 class (Class-10) clean room. Prior to mounting the individual sub-assemblies, planarity of the optical bench was ensured. For this, the mounting interfaces at the four corners of the optical bench were considered. Mounting interfaces at the four corners of the optical bench are EA, M1, grating, and 5303\,{\AA} channel detector mounting (Figure \ref{fig:1}).The planarity adjustment of the optical bench was carried out using a theodolite. The latter was fixed on a XYZ stage on top of a heavy duty tripod. A 100\,mm slip gauge was placed at the location of the EA mounting interface on the optical bench. Height and focus of the theodolite were adjusted such that the horizontal cross hair of the theodolite coincides with the top edge of the slip gauge. The reading of the theodolite was noted. The same experiment was repeated at the other three locations. In order to ensure the planarity of the optical bench, required shims (difference between values at EA, and other locations) were introduced in between the interfaces of the optical bench and the transfer trolley as per the values. The repeat of the experiment after fixing the shims indicated that the theodolite readings agree within $4^{\prime\prime}$.

\subsection{Optical Axis Reference Establishment}

After establishing the planarity of the optical bench, the theodolite axis was aligned with the optical axis of the payload. The optical axis of VELC needs to be at a distance of 195\,mm (design value) from the corner of the optical bench at the EA and M1 interfaces. The horizontal axis 
(H-axis) of the theodolite is adjusted to this with the help of slip gauges and theodolite. By design, the centers of all the optical components and the detector are at 130\,mm from the opto-mechanical interfaces at the top surface of optical bench. Therefore, the vertical axis of the theodolite, locations of the EA and the M1 were aligned to a height of 130\,mm from the mounting interfaces on the top surface of the optical bench using 130\,mm slip gauge. Height of the theodolite was adjusted such that its cross hair coincides with the tip of the height gauge placed at the EA and the M1 with vertical axis reading at $90^{\circ}$. The tilt error between the horizontal axis of the theodolite and the optical axis is 
${\approx}1.5^{\prime\prime}$, and ${\approx}2^{\prime\prime}$ between the vertical axis of the theodolite and the optical axis.

\subsection{ Primary Mirror (M1) Alignment}
 The M1 needs to be aligned interferometrically in its nominal position to ensure that no aberrations are introduced. Two theodolites were used to align the M1 mirror, theodolite-1 in front of the EA, and theodolite-2 in the opposite direction half a meter away from the M1 mirror. First, a reference flat was mounted at the M1 mounting interface and aligned with respect to theodolite-1 in auto-collimation mode. Then, a Zygo interferometer was aligned with respect to the optical axis of VELC using the reference flat. The position of the interferometer was fixed on the optical bench such that it is not disturbed during the entire process of aligning all the optical 
sub-assemblies on the optical bench. After aligning the interferometer, the actual flight mirror M1 was placed in its nominal position. Instead of the secondary mirror (M2), a flat mirror was mounted at the focal plane of M1. The flat mirror was adjusted such that the focused beam on it traces back the same path to the interferometer after reflection from the M1 through EA. The wavefront error in the reflected beam from the M1 was computed using the fringe pattern recorded by the interferometer. The RMS value was 8.34\,nm.  After aligning the M1 in its nominal position, the mounting screws were torqued to the optical bench while continuously monitoring the fringe pattern with the interferometer.

\subsection{Alignment of Secondary Mirror and the Collimating Lens Assembly}

For interferometric alignment of the M2 and the collimating lens, a custom-made spherical mirror with the same radius of curvature as the secondary mirror of VELC with RMS surface figure $\lambda$/40 was mounted at the secondary mirror location (M1 focus). A collimating lens assembly (CLA) was mounted in its nominal position in a free condition. With M2 (without central hole) and CLA in position, the wavefront error of the transmitted beam after the CLA was measured interferometrically using a reference flat with a RMS surface figure $\lambda$/100 kept after the CLA. The procedure similar to the alignment of the M1 mirror was repeated. The RMS error inferred from the interferogram 
was 16.8\,nm.

\subsection{ Alignment of the Quaternary Mirror (M4) and the Spectrograph Imaging Lens}

After aligning the collimating lens assembly (CLA), other components such as the M4, the Dichroic Beam Splitter-1 (DBS-1) and the spectrograph imaging lens were mounted on the optical bench at their respective nominal positions. The Reference flat with a RMS surface figure $\lambda$/100 was mounted at the focal plane of the spectrograph imaging lens. The tilt of the reference flat was adjusted such that the focused beam traverses back in the same path to the interferometer to measure the wavefront error. Following the procedure of the M1 mirror alignment, the wavefront error was found to be 13.7\,nm at the spectrograph multi-slit plane. The procedure was repeated after mounting the fold mirrors FM1, FM2, and FM3.

\subsection{Alignment of the Continuum channel}
Alignment of the continuum imaging lens and 5000\,{\AA} detector was done by measuring the point spread function (PSF) at the focal plane of the continuum channel. In the continuum channel 80\% of the energy was contained in 17.5\,${\mu}$m size (the design value was 16\,${\mu}$m).  After completion of the interferometric alignment of all the optical sub-assemblies, the secondary mirror was replaced with the actual flight mirror M2 with the hole. Then, the detectors were mounted and ground calibration of the instrument was performed, in both air and vacuum (\citealp{prasad2023}).

\subsection{Alignment of the Spectroscopic Channels}

The main optical assemblies of the spectroscopic channels in the VELC are Littrow lens assembly (LLA) and the diffraction grating. To align the LLA interferometrically, a flat mirror with a RMS surface figure $\lambda$/40 was used in place of the grating. The field lens and the LLA were mounted at their nominal positions. Orientation of the flat mirror was adjusted to generate the required angle of incidence. De-center and tilts of the field lens and the LLA were adjusted to minimize the wavefront error at the focal plane of the LLA. The interferometric maps showed 
  a RMS wavefront errors of 54.6\,nm and 54.7\,nm for the 5303\,{\AA} and the 7892\,{\AA} channels, respectively.

\section{Laboratory Calibration of the Instrument}

Generally, the different observing channels in the payload must be calibrated in the laboratory using a light beam whose charactetistics are similar to that in the regular observations in-orbit. In case of VELC, the instrument has a FoV up to 3\,$\mathrm{R}_\odot$ to study the solar corona with a cone angle of 1.6$^\circ$. The solar beam (i.e. sun disk) has a cone angle of 0.53$^\circ$. Feasibility of transmiting the sunlight inside the laboratory by using a coelostat on the rooftop and fold mirror was explored. But, this could not be realized in view of the nature of the laboratory building and considering the extreme clean conditions (Class 10) of the environment required for the payload to minimize the scatter in the instrument. Further, in spite of the laboratory measurements, instrument parameters need to be calibrated 
on board as these may change due to various factors while integrating with the spacecraft on ground, and in-orbit. We measured some of the parameters of the instruments in the laboratory at various stages of testing such as thermal cycle, vibrations, and other environment tests, both before and after integrating the payload with the spacecraft. We found that the parameters such as dark current, fixed pattern noise, response to light signal, stability of the images on the detectors, etc. remained similar and stable during all the stages of integration of the payload with the spacecraft.

\subsection{Wavelength Calibration Using Sunlight}

\begin{figure}[ht]
  \centering
  \includegraphics[width=0.99\textwidth,clip=]{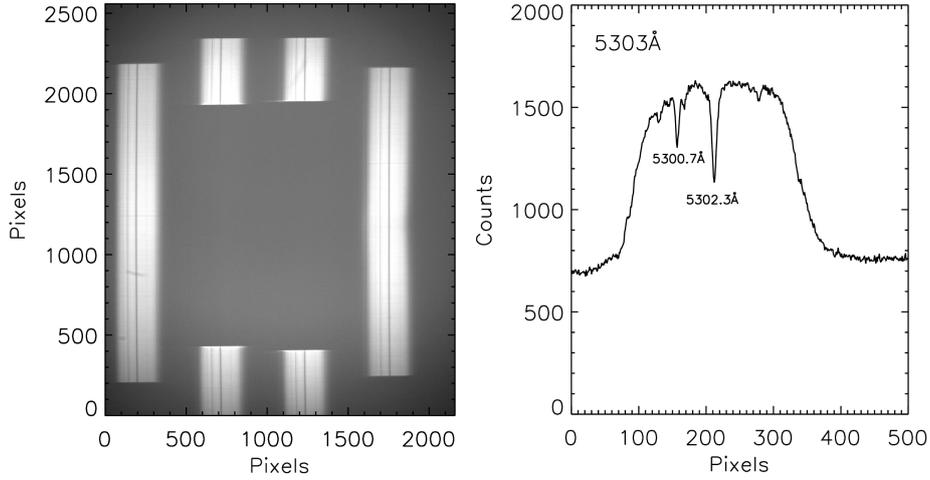} 
  \caption{The left panel shows the spectra around 5303\,{\AA} obtained with the sunlight transmitted inside the laboratory using a fibre bundle. The slits are 1, 2, 3 and 4 from the left to right. Unlikes slits 1 and 4, the middle region of slits 2 and 3 are not noticeable because of the occulter. The right panel shows the spectral profile near the center of  slit 1.}
  \label{fig:3}
\end{figure}

\begin{figure}[ht]
  \centering
  \includegraphics[width=0.99\textwidth,clip=]{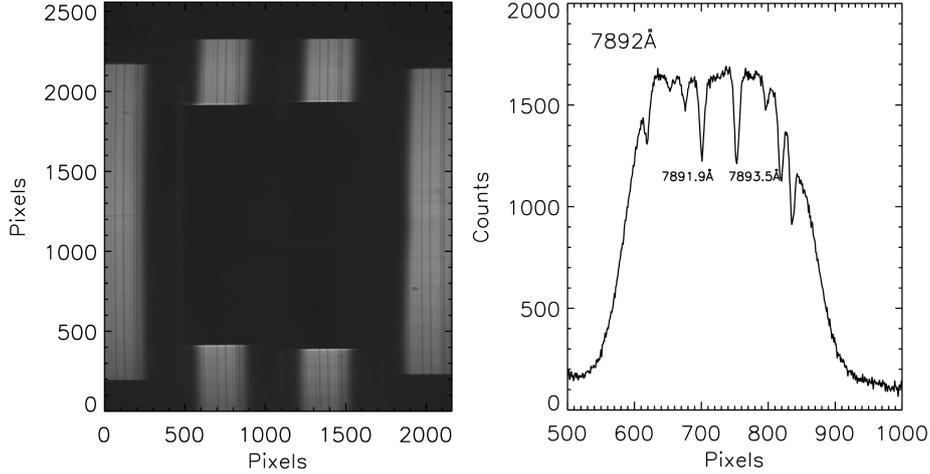} 
  \caption{Same as Figure 3 but for 7892\,{\AA}. The right panel shows the spectral profile near the center of slit 2 below the occulter in the left panel.}
  \label{fig:4}
\end{figure}

We were able to transmit the sunlight inside the laboratory using a 2\,mm fiber bundle. The light beam was expanded using a lens system, and directed at the slits of the spectrograph. The purpose of this experiment was to align the spectrograph and detectors to record the required part of the spectra. Figure~\ref{fig:3} shows the spectra of Sun for the four slits, and the spectral profile around the 5303\,{\AA} wavelength region,
in the left and right panels, respectively. 
The estimated dispersion from the separation between the two strong solar disk absorption lines at 5300.7\,{\AA} \& 5302.3\,{\AA} is
${\approx}$0.0284\,{\AA} per pixel. This is in good agreement with the design value $0.028$\,{\AA} (Table 1 and Section 3.6). The FWHM of the lines are 
${\approx}$0.25\,{\AA}. This is consistent with the expected value after taking into consideration the natural and instrumental broadening. 
The left and right panels of Figure~\ref{fig:4} shows the spectra and its profile for the 7892\,{\AA} wavelength region. The absorption lines have been marked in the spectral profiles. All the absorption lines in the 
7892\,{\AA} spectra are due to water vapor in the Earth’s atmosphere. The estimated dispersion from the separation between the two strong absorption lines at 7891.9\,{\AA} \& 7893.5\,{\AA} is
${\approx}$0.0313\,{\AA} per pixel. This is in good agreement with the design value $0.031$\,{\AA} (Table 1 and Section 3.6). The FWHM of the lines in this case is ${\approx}$0.30\,{\AA}. 
These observations confirm the alignment of detectors at the required wavelength region. The shape of the spectral profiles in the in-orbit observations are likely to be different due to different cone angle of the light beam, but the wavelength region will remain the same.
In the laboratory test, the light is a diverging solar beam due to the use of an expander lens assembly. The coronal light in-orbit will be a converging beam with cone angle of 1.6$^\circ$ passing through the narrow band filters. The spectra around 10747\,{\AA} could not be recorded due to cloudy sky conditions. The alignment of IR detectors was done using a tungsten lamp and narrow band filter. This procedure helps in seeing the spectrum due to each slit without absorption lines. The transmission curve of the filter was compared with that measured by the vendor to confirm the alignment.


\section{In-orbit Testing and Calibration}
\begin{figure}[ht]
  \centering
  \includegraphics[width=0.99\textwidth,clip=]{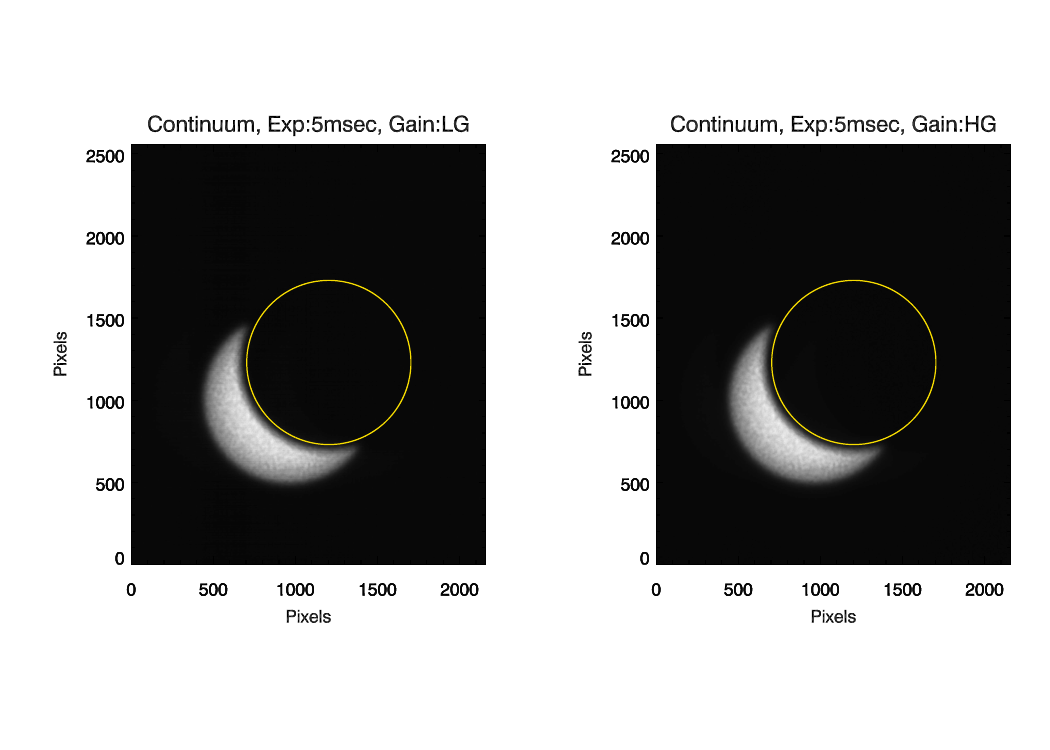} 
  \caption{The left and right panels show the continuum channel images of a portion of the Sun's disk protuding beyond the 
occulter, in low and high gain, when the spacecraft was pointed towards the centre of the Sun. The images show the offset between spacecraft and VELC reference axes. The black circular area  is the VELC occulter. The yellow open circle indicates the Sun disk size.}
  \label{fig:5}
\end{figure}

The spacecraft was launched on 2023 September 2, and injected in the `halo' orbit around the Sun-Earth Lagrangian L1 position on 2024 January 6. Some of the basic operations performed on the instrument have confirmed that all the components are in good aligned condition. The detectors, control units, and data acquisition systems are working well. There is some offset between the optic axis of the VELC and that of the spacecraft. Due to this, when the spacecraft is pointed towards the centre of the Sun, the image of the solar disk light does not fall exactly at the hole in the M2 miror (occultor). Hence, part of the solar disk image is formed on the detector.
The use of ND filter and its smaller aperture (45\,mm) helps in imaging the part of solar disk as seen in Figure \ref{fig:5} in the continuum channel. The exposure time was 5\,msec.

\subsection{Pointing of the Satellite}
 
After injection of the satellite into the `halo' orbit as mentioned in the previous paragraph, it was was pointed towards the Sun using the computed coordinates. The images in continuum showed that Sun center was off from the VELC occulter center by 
${\approx}7^{\prime}$, in both the roll and pitch directions (Figure \ref{fig:5}). Then, the satellite pointing coordinates were changed using the above computed values, and images in the continuum channel were obtained. The observations indicated that the Sun center was still off, but by a smaller margin. To align the center of the Sun with that of the occulter (hole in M2 mirror) accurately, the satellite pointing was changed by 
${\approx}1^{\prime}$, in steps of $10^{\prime\prime}$ in both the roll and pitch directions. The coordinates where there was symmetric illumination around the occulter were considered. In this position, the Sun and occulter centers coincide within an uncertainty of $10^{\prime\prime}$. Trial observations on several days showed that the corrections to the satellite pointing coordinates to align both the center of the Sun and occulter centers are the same, and stable. If any misalignment between the two centers is indicated by the observations in the future, a similar exercise will be carried out to point the satellite in the required direction.

\subsection{`Dark' Signal Measurements}

To carry out `dark' signal measurements on ground, we had installed the detector head in a thermally controlled vacuum tank in the laboratory. The glass windows were closed with thick layers of black paper and cloth. Measurements were obtained with different exposure times (\citealp{priyal2023}) at the temperature of $22^{\circ}$C, the expected ambient temperature of the payload during the regular observations in-orbit. `Dark' signal measurements 
in-orbit were obtained by off-pointing the satellite to 
15$^{\circ}$ from the Sun center, so that no sunlight enters the payload. The shutter in front of the VELC EA was kept closed (Section 3.1). Both the ground and in-orbit measurements agree well.

\subsection{Variation of the Signal with the Exposure Time for the Continuum Channel}
\begin{figure}[ht]
  \centering
  \includegraphics[width=0.99\textwidth,clip=]{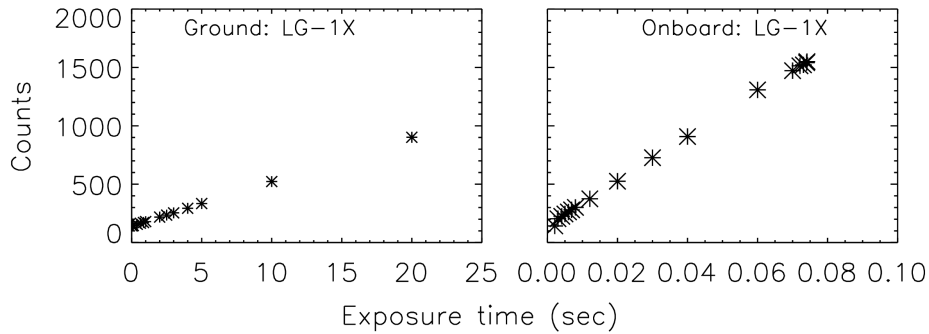} 
  \caption{Left and right panels show variations in the continuum channel detector mean count (light-dark) for 1X gain as a function of the exposure time for ground and in-orbit, respectively.}
  \label{fig:6}
\end{figure}

We find that the values of in-orbit `dark' counts for the detectors in VELC
are similar to those at the ground level. But the variation (fixed pattern noise) in the `dark' count appears to be different from that observed at the laboratory. This may be due to change in the electronics for data acquisition, and the harness cables used. 
The left and right panels in Figure~\ref{fig:6} shows the variation of in-orbit detector count with exposure time for 1X gain (using the solar image seen in Figure~\ref{fig:5}, but at different exposure times) and in the laboratory with a uniform light source, respectively. The exposure times are different in the two cases because of the nature of the light sources. The detector response w.r.t to the exposure times is linear in both cases.

\subsection{Identification of Hot/Dead Pixels}

In the case of CMOS detectors, neither the `dark' frames nor the images obtained with uniform light illumination showed any hot/dead pixels. None of the pixels indicated extreme or very low signal with exposure times in the range 0.1\,-\,100 sec. But in the case of the IR detector, a particular location always indicated extreme signal irrespective of the exposure time. The corresponding area expanded to the neighbouring pixels for larger exposure times (\citealp{priyal2023}).

\subsection{Data Quality Index}
 
The observed data are compressed in the binary format on board the satellite, transmitted to the ground station, decompressed, and then disseminated to the computer center at the Indian Institute of Astrophysics. We find that sometimes data in some rows of the image are not received in this process. Zeros are included for the missing rows/data. Depending on the number of missing rows, the images are labeled as either good or average or poor, and indicated in the header for the corresponding data files.

\subsection{Spectroscopic Observation}
\begin{figure}[ht]
  \centering
  \includegraphics[width=0.99\textwidth,clip=]{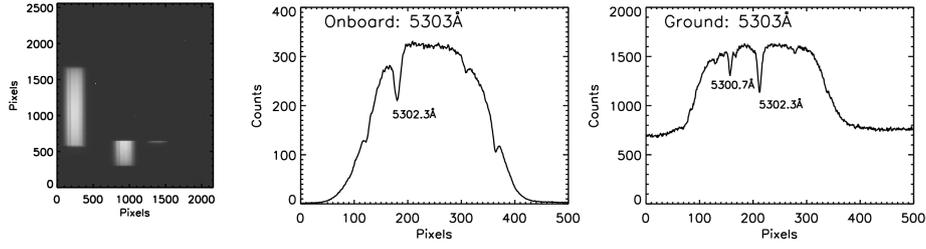} 
  \caption{Left panel shows the in-orbit spectrum of the solar disk around 
  5303\,{\AA} obtained through the ND filter in the EA shutter closed condition (Section 3.1). The sunlight was incident partially on slits 1  and 2 of the spectrograph (Figure \ref{fig:5}). The middle and right panels show typical profiles of the spectra of sunlight for the in-orbit and ground observations, respectively.}
  \label{fig:7}
\end{figure}

\begin{figure}[ht]
  \centering
  \includegraphics[width=0.99\textwidth,clip=]{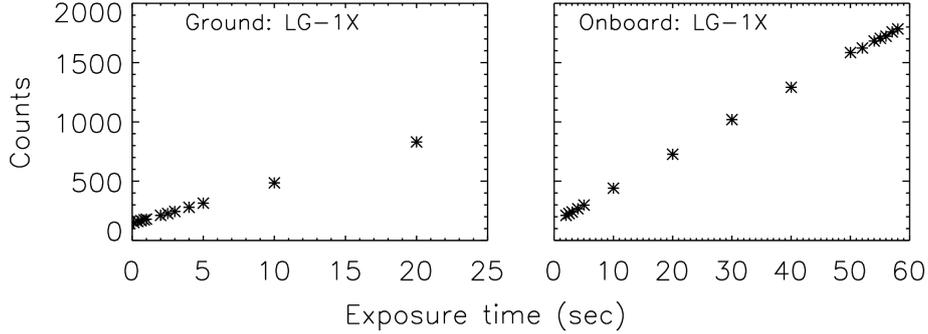} 
  \caption{Left and right panels show the count at the continuum portion of the spectrum around 5303\,{\AA} as a function of exposure time for laboratory and in-orbit observations, respectively. The X-axis and Y-axis scales are different due to difference in the nature of the light sources.}
  \label{fig:8}
\end{figure}

\begin{figure}[ht]
  \centering
  \includegraphics[width=0.99\textwidth,clip=]{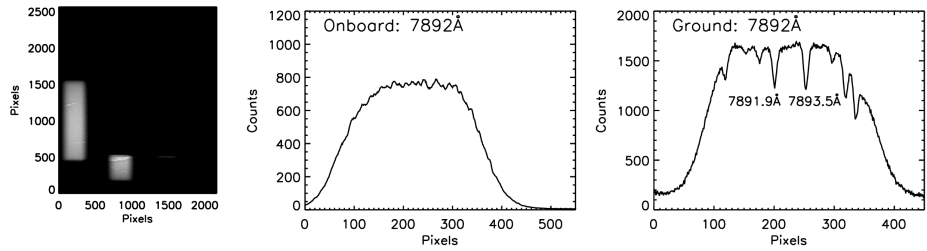} 
  \caption{Left panel of the figure shows the in-orbit spectrum of the solar disk around 7892\,{\AA} wavelength regions obtained through the ND filter in the EA shutter closed condition (Section 3.1). The sunlight was incident only on slits 1 and 2 of the spectrograph 
(Figure \ref{fig:5}). The middle and the right panels show typical profiles of the spectra of sunlight for the in-orbit and ground observations, respectively.}
  \label{fig:9}
\end{figure}

\begin{figure}[ht]
  \centering
  \includegraphics[width=0.99\textwidth,clip=]{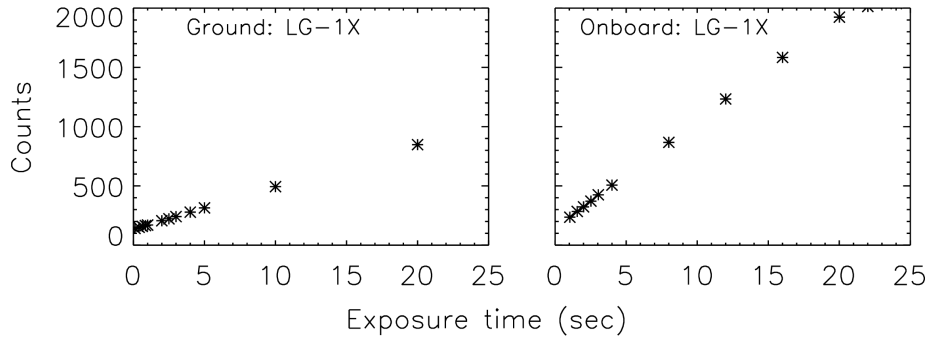} 
  \caption{Left and right panels of the figure show the count at the continuum portion of the spectrum around 7892\,{\AA} as a function of exposure time for ground and in-orbit observations, respectively. The X-axis and Y-axis scales are different due to difference in light source.}
  \label{fig:10}
\end{figure}

While obtaining images of part of the solar disk through the ND filter (Figure \ref{fig:5}), spectroscopic observations in all the three channels (5303\,{\AA}, 7892\,{\AA}, and 10747\,{\AA}) were also carried out. The left panel of Figure~\ref{fig:7} shows the spectra due to slit 1 and the lower portion of slit 2. The sunlight was incident partially on these slits 
(Figure \ref{fig:5}). The middle and right panel of Figure~\ref{fig:7} show profiles of the spectra obtained in-orbit and in the laboratory, respectively. The wavelengths of the identified absorption lines have been marked in each case. The transmission profiles of the filter appear different as the cone angle and angle of the beam with respect to the filter are different in the two cases. The transmission curve of the filter is expected to change further for the coronal observations as the cone angle of the beam passing through the narrow band filter will increase. The 5303\,{\AA} emission line will be adjacent to the absorption line at 5302.3\,{\AA}. The width of the emission line will be large, almost ten times that of the absorption line because of the high temperature of the coronal plasma leading to thermal broadening of emission lines. We chose a region of interest (ROI) in the continuum portion of the disk spectrum avoiding the absorption line. Same ROI was selected for the data with different exposure times. The left panel of Figure~\ref{fig:8} shows the signal as a function of exposure time for the low gain (1X) observations. For comparison we have plotted the signal  vs. the exposure time in the right panel for the same detector using data obtained in the laboratory with uniform light source. The detector response w.r.t to the exposure times is linear in both the cases. Similar analysis was carried out for the spectra obtained around the 
7892\,{\AA} line. Figure~\ref{fig:9} shows the spectra and profiles of the 
7892\,{\AA} wavelength region obtained 
in-orbit and the laboratory. The strong absorption lines observed during the tests in the laboratory are due to water vapour in the atmosphere. Hence, these absorption lines are not seen in the in-orbit spectrum. Left and right panels of the Figure \ref{fig:10} show the counts for the continuum portion of the spectrum as a function of exposure time for the laboratory and in-orbit observations at 1X gain. Observations at other gains of the detectors show expected performance. The in-orbit spectrum around 10747\,{\AA} was noisy and difficult to analyze. 

\subsubsection{Raster Scan Observations}
\begin{figure}[!htbp]
  \centering
  \subfloat[]{\includegraphics[width=0.69\textwidth,clip=]{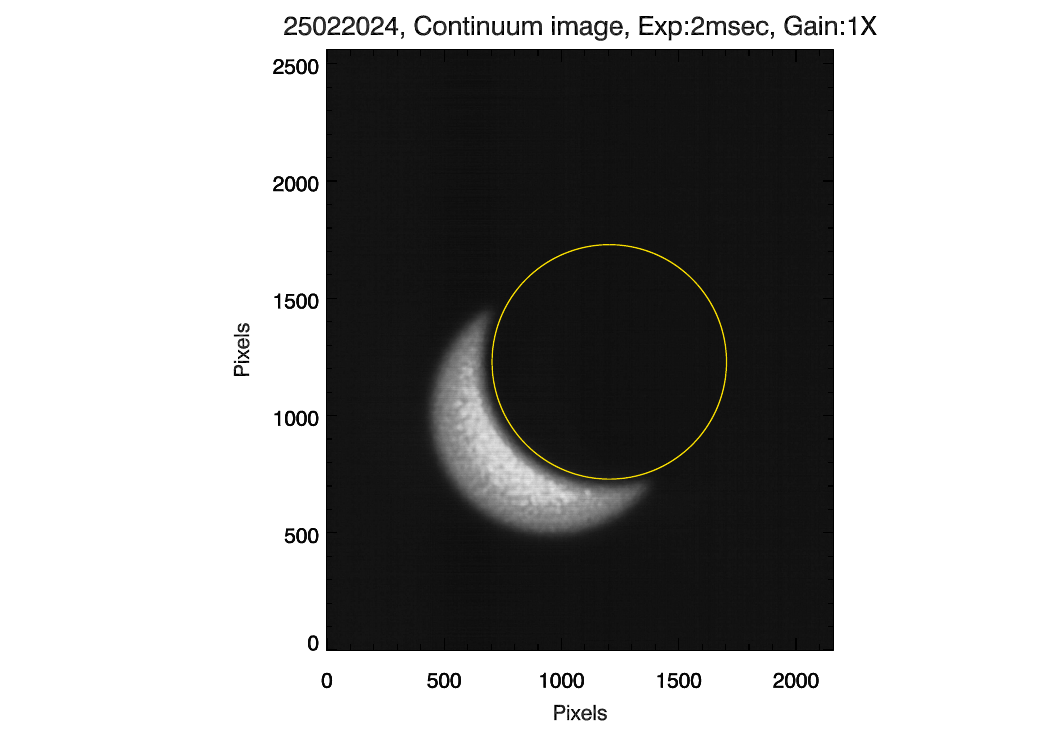} \label{fig:11a}} \\
  \subfloat[]{\includegraphics[width=0.69\textwidth,clip=]{fig_11b.pdf} \label{fig:11b}}
  \caption{Upper and lower panel images show the continuum image (shown in Figure \ref{fig:5}), and the raster scan image constructed from 5303\,{\AA} observations, respectively. The black circular area in both the images is the VELC occulter. The yellow open circle indicates the Sun disk size. The missing portion in the lower left corner of the raster scan image is due to the FoV of slit 1.} \label{fig:11}
\end{figure}

We carried out raster scan observations to confirm the working of LSM. 
The spectra in 5303\,{\AA}, 7892\,{\AA} and 10747\,{\AA} (IR) channels were obtained with exposure times of 4\,sec, 2\,sec, 20\,sec respectively, with 1X and 10X gains simultaneously. The LSM was moved from 
-940\,${\mu}$m to +940\,${\mu}$m in steps of 20\,${\mu}$m (Section 3.5). Generally parameters such as central wavelength of the line, its intensity and width need to be computed to study the spatial and temporal variations. The solar corona cannot be observed in the EA shutter closed condition (Section 3.1). So, we decided to generate an intensity image of  part of the solar disk (Figure \ref{fig:5}) from the spectroscopic observations and compare it with the corresponding image obtained in the continuum channel. For this purpose, we considered the average intensity of six pixels in the spatial direction, at the specified wavelength in the continuum. This was uniformly followed for all the locations along the length of the slit (spatial direction) and for all the LSM positions in the raster scan image. Six pixels average was considered for 5303\,{\AA} and 7892\,{\AA} channels since their size (6${\times}$6.5\,${\mu}$m) is close to the LSM step size of 20\,${\mu}$m and the equivalent 
40\,${\mu}$m shift of the image at the slit plane (Section 3.5). Figure \ref{fig:11}a shows part of the solar image taken in the continuum channel, and Figure \ref{fig:11}b shows the image constructed from the raster scan observations. 
The raster scan image confirms the smooth functioning of the LSM.  Note that the continuum and raster scan images have different spatial scales, thus appear different in size.

\subsection{Conversion of Count Measurements to Standard Units}

Solar disk measurements at the continuum portion of the spectrum can be used as a standard source as it varies by ${\approx}$0.2\% only over a solar cycle. Therefore, we obtained the disk spectra via the 45 mm aperture neutral density filter (ND) when the EA shutter was closed (Section 3.1) and satellite was 
off-pointed by 
$16^{\prime}$ from the Sun center. The ND filter has a transmission of 
$0.64{\times}10^{-4}$ at 5303\,{\AA}. We find that average count per pixel is 
${\approx}$30 in the continuum portion of the spectrum adjacent to 5303\,{\AA} emission line. The above count is equivalent to ${\approx}176{\times}10^{6}$ when the 147\,mm EA is open and the ND filter is not present in the path of the light beam. The solar flux at the center of disk image is 
$3.0458{\times}10^{6}$ ergs\,s$^{-1}$cm$^{-2}${\AA}$^{-1}$sr$^{-1}$ at 5303\,{\AA}. Hence, one count is equal to 0.0173 ergs\,s$^{-1}$cm$^{-2}${\AA}$^{-1}$sr$^{-1}$. The conversion factor will be verified once every three months by carrying out similar observations. 
The observed peak intensity of the emission line in the corona varies between 15 and 100 at 1.1\,$R_{\odot}$ for the bright regions. Considering the average peak intensity of 57.5 counts for the emission line in the observed coronal region and a typical line-width of 0.9\,{\AA}, the total intensity of the emission line 
is\,=\,57.5$\times$(0.9/0.0284)$\times$0.0173\,=\,32 ergs\,s$^{-1}$cm$^{-2}${\AA}$^{-1}$ sr$^{-1}$. This agrees well with the value of 
40 ergs\,s$^{-1}$cm$^{-2}${\AA}$^{-1}$ sr$^{-1}$ reported by \cite{koutchmy1983,singh2019}.

\section{Coronal Observations }
Considering the offset between the satellite and VELC axis, the satellite was off-pointed such that the VELC optical axis points towards the center of the Sun and the image of the latter lies at the center of the hole in the M2 mirror (occulter). Then, we carried out observations in the continuum channel to image the solar corona, and spectroscopic channels around three emission lines. The data in  the continuum, 7892\,{\AA} and 
10747\,{\AA} channels are noisy. Hence we report observations in the 5303\,{\AA} channel alone.

\subsection{Sit and Stare Observations}
It is possible to locate the image of the corona at the desired position with respect to the slit(s) by moving the LSM. By carrying out observations at various position of LSM from +940\,${\mu}$m to -940\,${\mu}$m, we found that 
the east edge of the occulter is at slit 1 when the LSM is at +160\,${\mu}$m position. Likewise, for -60\,${\mu}$m position of the LSM the west edge of the occulter is at slit 4 (refer Figure 3 for the slit numbers). At LSM position of +110\,${\mu}$m, the positions of the slits are symmetric with respect to Sun in the east-west direction. Slits 1 and 4 lie on the east and west side of the solar image, respectively. We have carried out observations at different cadences in the range 5\,-\,90\,sec. The high cadence observations are carried out for short duration to study periodic oscillations in the spectra of coronal structures, thereby the existence of waves, if present. The low cadence (60\,sec) observations will be used to study the dynamics of CMEs. 

\subsubsection{Flat-Fielding the Spectrum}

We have limitations in obtaining flat-field data for the continuum and spectroscopic channels. Since clean room conditions of Class 10 are to be maintained to minimize  scattering inside the instrument during the laboratory tests, it was very difficult to create light beam of size and f-ratio similar to that of the expected coronal beam in the regular in-orbit observations. So, we obtained flat-field images in all the four channels with a uniform source created using Labsphere. 
Thus, these images record the fixed pattern noise of the detector only, and not intensity vignetting due to optics. We found that the in-orbit fixed pattern noise  differs from that obtained in the laboratory (Section 6.3). 

Generally, it is advisable to keep the narrow band filter close to entrance slit of the spectrograph. In this position, a small portion of the filter in width (dispersion direction) will be used. But, due to observational requirements in different wavelengths, we have kept the narrow band filters close to the detectors in each channel. This way, a large width of the filter about 3\,mm (transmission width of the filter) is used to record the spectrum. Different portions of the filter have different transmission, 
${\approx}$ 1\,-\,2\%. Also, the transmission profile of the filter varies with the angle of the incident beam. These can also cause some pattern in the recorded spectrum. Therefore, the use of the average transmission profile creates a large amount of uncertainty in the spectrum.  We have attempted to correct each spectrum for the transmission curve of the filter after performing the flat-field correction. It works well for the nearly flat portion of the transmission curve of the filter at the middle, but leaves significant differences for edge portions of the transmission curve. Therefore, we have considered only the central portion of the spectrum as seen in the middle panel of Figure \ref{fig:14} to derive the various parameters of emission line.

The use of flat-field images obtained in the laboratory for the in-orbit data improves the quality of image but not sufficient to accurately study the emission profiles. We found that fixed pattern noise for the in-orbit observations differs from that obtained in the laboratory. Futhermore, the in-orbit images show granular pattern (Figures \ref{fig:11} upper panel).
We have attempted to remove the fixed pattern noise in the in-orbit observations by using the spectral images obtained for LSM position at 
+940\,${\mu}$m and -940\,${\mu}$m where the emission at 5303\,{\AA} is negligible. Then, we constructed a flat-field image by interpolating the data for locations at the absorption lines, one for the data obtained for the LSM positions between 0\,${\mu}$m and +940\,${\mu}$m, and other for 
0\,${\mu}$m and -940\,${\mu}$m. The use of this flat-field image improves the quality of the spectral image significantly. It works well with the data obtained at different days, even though this is not the standard procedure to perform the flat-field correction.

\subsubsection{Stability of the Spectra in Dispersion Direction}

 Generally, spectra obtained with ground based observations do not shift on the detector in the dispersion and spatial directions due to the stability of the spectrograph. Our 
time sequence observations at fixed locations of the image show that spectral images are not stable in the dispersion (wavelength) direction. We found that the minimum in the absorption line at 5302.3\,{\AA} oscillates around the mean position by about 5 pixels ($\sim$140\,m{\AA}). This may be due to oscillations in some optical component or the detector. It is not clear why and how it happens.
We observed for number of days and concluded that the shift in the spectrum on the detector is not due to Doppler velocity of the plasma, but due to the instrument. Therefore, we have aligned all the spectra to the mean position of the absorption line adjacent to the 5303\,{\AA} emission line in the first slit (pixel no. 181, X-axis). After alignment, the variations in the center of the absorption line is within ${\pm}$1 pixel. 

\subsubsection{Curvature Correction for the Spectra}
Generally, the observed spectrum can be curved because of spherical aberrations in various optical components, especially in the short focal length spectrograph. The VELC spectra obtained showed curvature in the absorption line at 5302.3\,{\AA} even though the entrance slit is  straight. The difference in the location of absorption line is  
${\approx}$8 pixels between the edges and centre of the slit. We determined the center of the absorption line at each spatial location along the slit. A polynomial fit was computed for the data corresponding to the center of the absorption line. Then, the spectrum at each spatial location was shifted by the difference observed between the reference position of the absorption line and the polynomial fit value. After this, the error in the center of the absorption line was found to be  ${\pm}$1 pixel (28.4\,m{\AA} for the 
5302.3\,{\AA} line) at random, indicating a good alignment.

\subsubsection{Effect of Satellite Drift}
\begin{figure}[ht]
  \centering
  \includegraphics[width=0.99\textwidth,clip=]{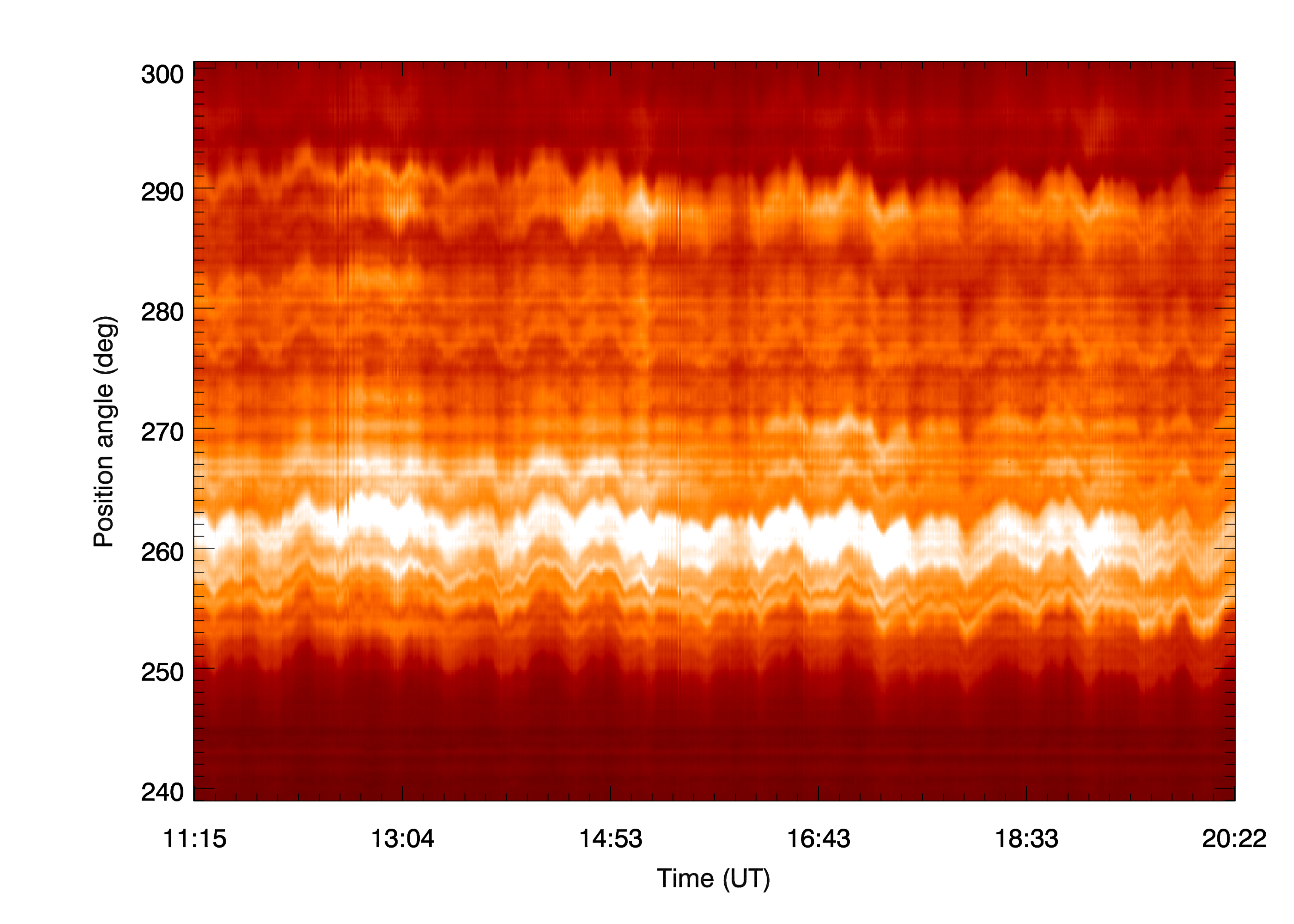} 
  \caption{Sit and stare observations of coronal intensity at 5303\,{\AA} on 2024 August 22. 
X-axis and Y-axis represents time and position angle, respectively. Variations in the intensity between different position angles are due to the brightness of the coronal structures at different locations. The oscillatory pattern along the X-axis may be due to the drift of the satellite or some unknown factor.}
  \label{fig:12}
\end{figure}

We also found that the coronal spectra show an oscillatory pattern along the slit (spatial direction). We obtained spectra with an exposure time of 
5\,sec and cadence of 50\,sec by keeping the LSM at -60\,${\mu}$m for about 9\,h on 2024 August 22. Figure \ref{fig:12} shows the intensity at various locations (i.e. position angles, measured counter clockwise from the solar north to east) along slit 4 with time. It is a long duration data set obtained at low cadence. The oscillatory pattern as a function of time is clearly evident. This may be due to satellite pointing or some unknown reason. The shift is 
${\approx}{\pm}$20$^{\prime\prime}$ about the mean position.  
We attempted to compensate for this drift along the slit using the brightest feature seen in Figure \ref{fig:12} near position angle ${\approx} 260^{\circ}$.  We selected consecutive 20\,-\,40 intensity locations along the spatial direction around the bright feature, and determined the location of the maximum intensity in each spectrum of the time sequence observations. Then we computed the mean spatial location of the maximum intensity. Assuming the location of the maximum intensity does not vary much with time, we align all the spatial locations at each time sequence in the observations to the above mentioned mean position. 
The work is in progress and will be reported separately.
Note that in the absence of a bright feature, there can be uncertainities in determining the location of the maximum intensity.

\subsubsection{Estimate of Scatter Light}

It is difficult to estimate the scatter in the instrument as coronal and scattered light are always present in the continuum and spectral images. An indirect approach may help to get an estimate of the scattered light in the spectrograph. We assume that the contribution of the continuum coronal light is about 2\% of the emission light in the 5303\,{\AA} line (equivalent width of emission line\,=\,60), and can be neglected. Therefore, the continuum signal in the coronal spectra is mostly because of scattered light in the instrument. We found in Section 6.7 that solar disk count is 
$176{\times}10^{6}$ pixel$^{-1}$s$^{-1}$ for 1X gain of the detector in the EA shutter open observations 
(147\,mm EA, without the ND filter). The average count at the continuum portion of the coronal spectrum is 
${\approx}$200 pixel$^{-1}$s$^{-1}$ for 1X gain, for the slit position at 
1.13\,$\mathrm{R}_\odot$. It may be noted that there are variations in the above count for different days of observations. The ratio of continuum signal to the disk light in the coronal spectra is ${\approx}1.1{\times}10^{-6}$. The ratio is expected to be much higher close to the solar limb since scattered light decreases exponentially. Note that the emission corona signal at 
1.13\,$\mathrm{R}_\odot$ is ${\sim}10^{-8}$. It varies depending on the coronal structures (Section 6.7).


\subsubsection{Data Analysis}
We have analysed the data obtained at  a number of LSM positions with different gains and exposure times. We find that spectral images with 
5\,sec exposure in low gain mode (1X) are of comparatively good quality as compared to other exposure times. Exposure times $>$5\,sec leads to saturation of the spectra near the solar limb. Adding the spectra on board to increase the SNR has a drawback. The spectra obtained by adding the frames on board get averaged in the dispersion and spatial directions as the 
spectrum shows oscillations in both these directions (Sections 7.1.2 and 7.1.4).

\begin{figure}[]
  \centering
  \includegraphics[width=0.99\textwidth,clip=]{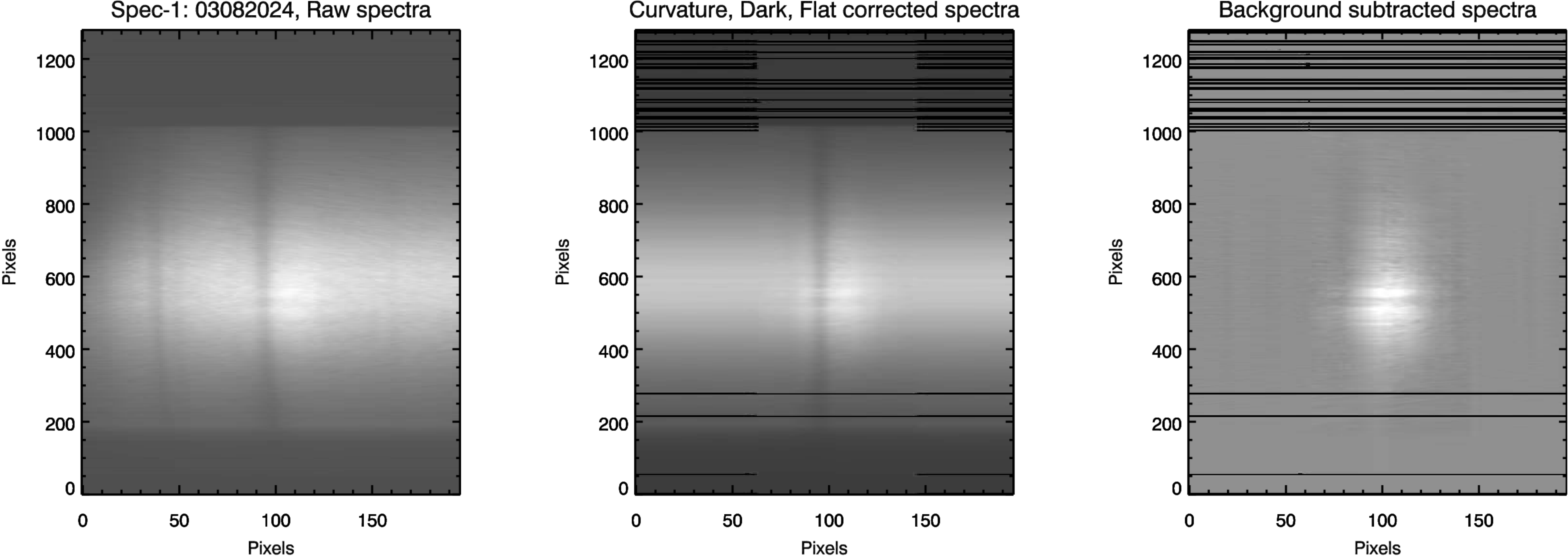} 
  \caption{ The left panel shows the spectrum around 5303\,{\AA} for 1X gain obtained with 5\,sec exposure time on 2024 August 3. X- and Y- axes represent dispersion and spatial directions, respectively.The middle panel shows the spectrum after curvature, dark, and flat-field corrections.
  The right side panel shows the spectrum after background subtraction.}
  \label{fig:13}
\end{figure}

The left side panel of Figure~\ref{fig:13} shows a typical spectrum obtained with 5 sec exposure at LSM at +160\,${\mu}$m in sit and stare mode of observations. The middle panel shows the spectrum after curvature correction, dark subtraction and flat-field correction. The right side panel shows the spectrum after background subtraction. We notice that the amplitude of signal due to emission line at 5303\,{\AA} is about 10\%\,-\,15\% whereas background count is 85\%\,-\,90\% in these observations. This is the case in most of the observations. For observations made by \cite{singh2003a} with 25\,cm coronagraph at Norikura observatory, the emission line signal and the background are about 60\% and 40\%, respectively. 
It is not clear why the corresponding values are siginficantly different in VELC observations. Probably, some stray light is entering the instrument. This assumption gains support from the observations of the deep sky in continuum channel. The satellite was off pointed by about 75$^{\circ}$, looking far away from the Sun and images of Sirius A (stellar observation) were obtained in the continuum channel. 
The detector showed high background count, indicating the existence of stray light. 

\begin{figure}[!htbp]
  \centering
  \includegraphics[width=0.99\textwidth,clip=]{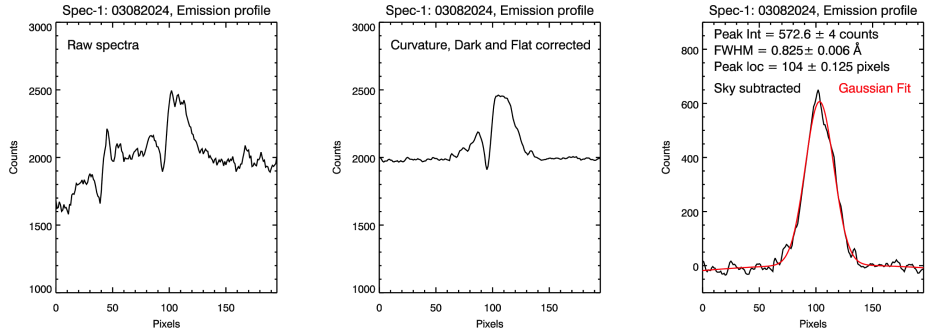} 
  \caption{The left panel shows the profile around 5303\,{\AA} for 1X gain obtained with 5\,sec exposure time on 2024 August 3 at a spatial location (Figure \ref{fig:13}).The middle panel shows the profile for the spectrum shown in the middle panel of Figure \ref{fig:13} after curvature, dark, and flat-field corrections. The right side panel shows the profile after background subtraction along with a Gaussian fit. The parameters along with the SD values of the Gaussian fitted profile are given in panel.}
  \label{fig:14}
\end{figure}

Left panel of Figure \ref{fig:14} indicates the observed profile around 
5303\,{\AA} at a spatial location for the spectra shown in the left panel of Figure \ref{fig:13}. Middle and right panels of Figure \ref{fig:14} show the profiles for the spectra shown in middle and right panels of Figure \ref{fig:13}, respectively. The red colour profile in the right side panel is the Gaussian fit for the corrected profile. The parameters such as intensity, FWHM, and Doppler velocity of the line profile have been computed from the Gaussian fit, and stored as a function of spatial location and time. Figure 15a and b show the histograms of FWHM for the data of slit 1 (east) and slit 4 (west) respectively. The values of FWHM for both the slits 
are in the expected range 0.5\,-\,1.5\,{\AA} considering the temperature 
(1\,-\,3${\times}$10$^{6}$\,K) and non-thermal velocity for microscale 
Fe{\sc xiv} emission. The values of FWHM have been corrected for the instrumental line width. The difference between the mean values of the FWHM for the solar corona at the east and west edges of the occulter may be due to difference in the physical characteristics of the coronal structures at either location on that day. The larger width of the distribution in the case of slit 1 (east) could be due to contribution from different loops (behind and front) also. We plan to investigate this using larger data set, obtained on different days.


\begin{figure}[!htbp]
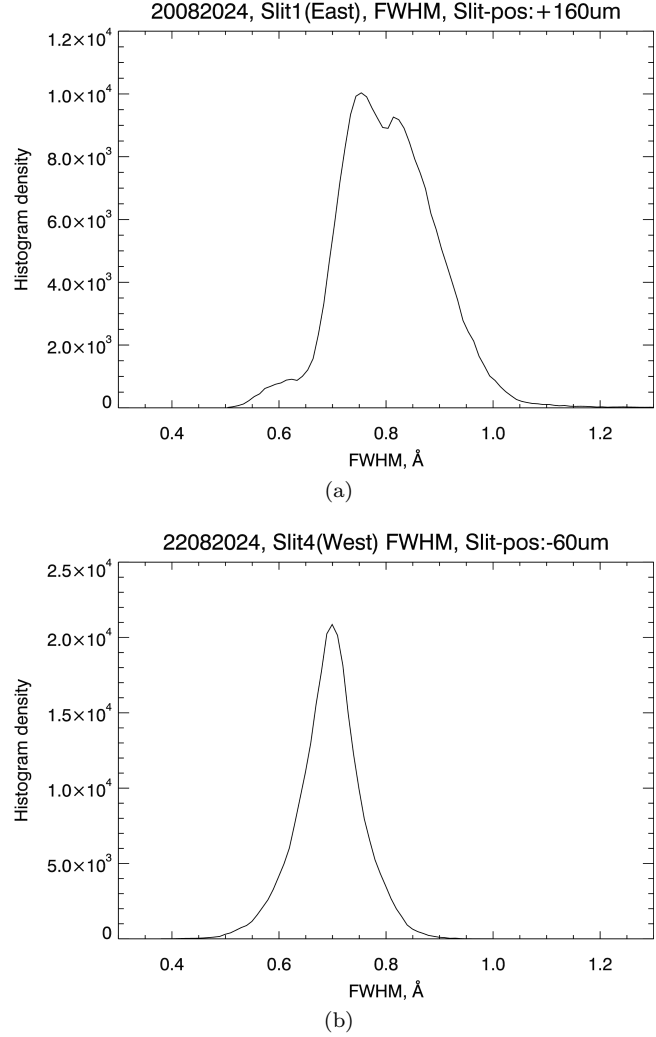

  \centering
  \subfloat[]{\includegraphics[width=0.69\textwidth,clip=]{fig_15a.pdf} \label{fig:15a}} \\
  \subfloat[]{\includegraphics[width=0.69\textwidth,clip=]{fig_15b.pdf} \label{fig:15b}}
  \caption{The upper and lower panels show the histograms of FWHM for the data of slit 1 (east) and slit 4 (west), respectively.} \label{fig:15}
\end{figure}


\subsection{Raster Scan Observations: Coronal}
\begin{figure}[!htbp]
  \centering
   \includegraphics[width=0.99\textwidth,clip=]{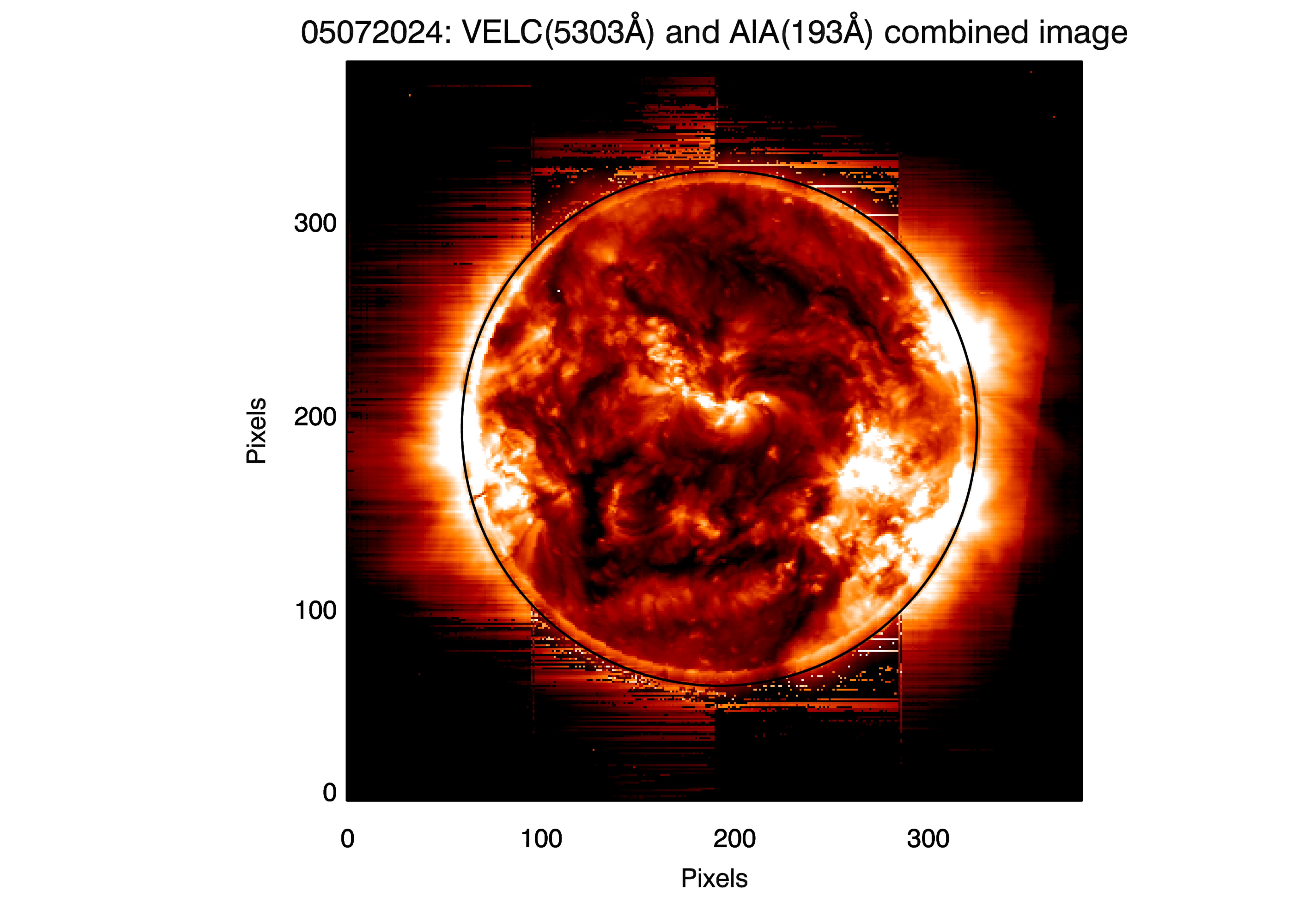}
  \caption{Composite intensity image of the solar corona in 5303\,{\AA} (VELC) and 193\,{\AA} (SDO/AIA) on 2024 July 5. The black circle represents the VELC occulter.}
  \label{fig:16}
\end{figure}

We carried out raster scan observations with different combinations of exposure times, on board frame binning, different gains of the cameras, and LSM step sizes in all the spectroscopic channels. Considering the spectral images, time required to complete one raster scan, data rate, and data volume per raster scan, we find that the optimum parameters for raster scan observations are:
exposure time\,=\,4\,sec; number of frames binned on board (i.e. frame binning count, FBC)\,=\,5; camera gains\,=\,1X and 10X; step size\,=\,20\,${\mu}$m; number of 
steps\,=\,95 (-940\,${\mu}$m to +940\,${\mu}$m); time for one raster 
scan\,=\,32\,min. Note that raster scans can be taken with different parameters to get the data at smaller or larger cadence depending upon the scientific objective. Data volume should be also considered while planning such observations. Here, we present the results of the raster scan made with the above mentioned parameters. We have analyzed the raster scan data using all the steps mentioned earlier for sit and stare mode observations. We computed the parameters such as the central wavelength of the emission line, intensity and width of emission line at each spatial location along the slit and for each step of the LSM. The peak intensities were determined in counts as shown in the right side panel of 
Figure \ref{fig:14}. The counts can be converted in to solar flux units considering number of pixels binned and exposure time used during observations as explained in Section 6.7. There are four slits and we analyzed the data for each slit separately. After determining the parameters of emission line, we combined the data from all the four slits to form a coronal image. Figure \ref{fig:16} shows a composite of the intensity images of the solar corona observed with VELC in 5303\,{\AA} (FoV${\approx}$1.5\,$\mathrm{R}_\odot$) and 
SDO/AIA in 193\,{\AA} (FoV${\approx}$1.2\,$\mathrm{R}_\odot$). 
There is a good correspondence between the off-limb structures in the two images. 
The coronal features above the east and west limbs of the Sun could be noticed in the VELC image. Due to the larger FoV in the VELC, the coronal loops (particularly those above the west limb) could be observed to comparitively larger distances. But no coronal structures could be noticed above the north and south limbs of the Sun. The emission is very weak there. 
Computing the intensity from emission line profile for weak coronal features is difficult amidst the background continuum. Gaussian fit for the line profile does not converge for weak features and hence we do not notice the latter in both the polar regions. It is to be noted that there are no bright coronal structures near the north and solar limbs in the SDO/AIA image also.

\section{Summary}

We find that the background signal in the coronal observations with VELC is large, more than expected. Spectroscopic observations in the
5303\,{\AA} emission line with the 25\,cm ground based Norikura coronagraph indicate a large ratio (${\approx}$ 1.5) of emission signal to the background nearby continuum portion of the spectrum as compared to the ratio of 
${\approx}$0.3 in the VELC observations. This indicates more scatter light in VELC.
Larger than design microroughness of the M1 mirror is likely to increase the scattering, but not certain if that is the only cause. 
The observations in the 5303\,{\AA} line are useful in the sit and stare mode as well as raster scan mode (\citealp{ramesh2024}). The authors reported direct proof of coronal dimming during the occurrence of a CME. The sit and stare mode with high cadence observations will help to study oscillations in the coronal structures, thereby existence of waves, flows and dynamics of corona. The sit and stare mode observations with low cadence for long duration will help to study the CMEs \citep{priyal2025}, and other energetic events along with observations in the radio wavelengths \citep{ramesh1998,ramesh2012,kathiravan2002,hariharan2014}. The raster scan observations will help to study the variation in emission line parameters spatially, thereby the physical characteristics of coronal structures. Raster scan observations will also permit to investigate the long term variations in the solar corona. Joint observations with VELC and instruments like AIA, uCoMP, PROBA-3 are expected to be a valuable tool to understand the near-Sun corona.

The spectroscopic observations carried out in the 7892\,{\AA} channel with different exposure times, and adding number of spectra to enhance the signal to noise ratio has not resulted in the detection of the emission component at 7892\,{\AA} wavelength. We tried to do this in all the gains of the detector, but the emission line signal was elusive. Also, emission signal at 7892 {\AA} line is less as compared to 5303\,{\AA}. The use of narrow band filter introduces noise in the data (Section 7.1.1). The ground based observations indicated that data with IR detector has large noise, both in low and high gain mode of operation. The noise in the in-orbit data has increased  further, probably due to EMI. Repeated experiments for 
10747\,{\AA} channel, similar to 7892\,{\AA} channel, has not yielded positive results. At present, it appears that the use of data from these channels  for scientific purposes might be difficult. Attempts may be made in the future to retrieve useful data from these two channels. Some of the scientific objectives, for example temperature determination from the intensity ratios of two emission lines, magnetic field measurements, etc. will not be possible till then. In spite of these limitations, continuous observation in the 5303\,{\AA} channel will be very valuable to study the corona, especially energetic events.

\section*{Acknowledgments}
Aditya-L1 is an observatory class mission which is fully funded and operated by the Indian Space Research Organization (ISRO).
We thank all the scientists and engineers at the various centres of ISRO and Indian Institute of Astrophysics who have made signficant contributions for the VELC payload to reach the present state. 
The SDO team is acknowledged for online dissemination of their data. We thank the referee for his/her valuable comments which helped us to present the results more clearly. 

\section*{Declarations}
No funding was received for conducting this study.



\end{document}